\documentclass[aps,twocolumn,showpacs,pra,superscriptaddress]{revtex4-2}
\usepackage{mathtools}
\usepackage{bm}
\usepackage{dsfont,amsthm,amsbsy}
\usepackage{verbatim}
\usepackage{amssymb}
\usepackage{amsmath}
\usepackage{bbm}
\usepackage{graphicx}
\usepackage{epstopdf}
\usepackage{subfigure}
\usepackage{natbib}
\usepackage{epsfig}
\usepackage{amsfonts}
\usepackage{mathrsfs}
\usepackage{sidecap}
\usepackage{lipsum}
\usepackage{url}
\usepackage[toc,page,title,titletoc,header]{appendix}
\usepackage[colorlinks,linkcolor=blue,citecolor=blue,anchorcolor=blue,urlcolor=blue]{hyperref}
\usepackage{hyperref}
\usepackage{resizegather}
\usepackage{tikz}
\usepackage{float}
\usepackage{mathbbol}
\usepackage[normalem]{ulem}
\usepackage{cancel}
\usepackage{upgreek}
\usepackage{enumitem}

\newcommand{\bl}{\begin{aligned}}
\newcommand{\el}{\end{aligned}}

\def\be{\begin{equation}}
\def\ee{\end{equation}}

\def\bi{\begin{itemize}}
\def\ei{\end{itemize}}
\def\bn{\begin{enumerate}}
\def\en{\end{enumerate}}
\def\bea{\begin{eqnarray}}
\def\eea{\end{eqnarray}}
\def\no{\nonumber}
\def\ba{\begin{array}}
\def\ea{\end{array}}
\def\bd{\begin{displaymath}}
\def\ed{\end{displaymath}}

\begin{document}

\title{Anti Kibble-Zurek behavior in the quantum $XY$ spin-$1/2$ chain driven by correlated noisy magnetic field and anisotropy}

\author{S. Sadeghizade}
\email[]{saj.sadeghi@sharif.edu}
\affiliation{Department of Physics, Sharif University of Technology, 11155-9161, Tehran, Iran}

\author{R. Jafari}
\email[]{jafari@iasbs.ac.ir, raadmehr.jafari@gmail.com}
\affiliation{Department of Physics, Institute for Advanced Studies in Basic Sciences (IASBS), Zanjan 45137-66731, Iran}
\affiliation{School of Quantum Physics and Matter, Institute for Research in Fundamental Sciences (IPM), Tehran 19538-33511, Iran}
\affiliation{Department of Physics, University of Gothenburg, SE 412 96 Gothenburg, Sweden}

\author{A. Langari}
\email[]{langari@sharif.edu}
\affiliation{Department of Physics, Sharif University of Technology, 11155-9161, Tehran, Iran}

\begin{abstract}
In the non-adiabatic dynamics across a quantum phase transition, the Kibble-Zurek paradigm describes that the average 
number of topological defects is suppressed as a universal power law with the quench time scale. 
A conflicting observation, which termed anti-Kibble-Zurek dynamics has been reported in several studies, specifically 
in the driven systems with an uncorrelated stochastic (white) noise.  
Here, we study the defect generation in the driven transverse field/anisotropy quantum $XY$ model in the presence of a correlated (colored) Gaussian noise.
We propose a generic conjecture that properly capture the noise-induced excitation features, which shows good agreement with the numerical simulations.
We show that, the dynamical features of defect density are modified by varying the noise correlation time.
Our numerical simulations confirm that, for fast noises, the dynamics of the defect density is the same as 
that of the uncorrelated (white) noise, as is expected. However, the larger ratio of noise correlation time to the annealing time results in larger defects density formation and reforms the universal dynamical features.
Our finding reveals that, the noise-induced defects scale linearly with the annealing time for fast noises, while in the presence of the slow noises, 
the noise-induced defects scale linearly with the square of the annealing time.  
The numerical simulations confirm that, the optimal annealing time, at which the defects density is minimum, scales linearly in logarithmic scale with the total noise power having different exponents for the fast and slow noises.

\end{abstract}

\maketitle

\section{\label{sec:level1}Introduction\\}

The Kibble-Zurek mechanism (KZM) provides a predominant theoretical framework for exploring the critical dynamics
of phase transitions in systems ranging from cosmology to condensed matter physics \cite{kibble_topology, zurek_1996}.
According to KZM, in thermodynamic limit, no matter how slowly a system is driven across the transition point, its evolution cannot be adiabatic 
close to the critical point (gap closing point) \cite{dziarmaga_2010}. Therefore, after crossing the transition point the system is excited 
and topological defects are created in the system. Density of these defects follows a universal power law and is proportional to 
$\tau_{Q}^{-d \nu/(1 + \nu z)}$, where $\tau_Q$ is the quench time scale, $d$ is the dimensionality of the system, $\nu$ and $z$ 
are correlation length and dynamical critical exponents of the system, respectively \cite{dziarmaga_2010,zurek_dynamics_2005,dutta_akzm}.

KZM for classical continuous phase transitions has been verified for many different systems \cite{navon_2015,ulm_2013,pyka_2013,du_2023,lee_2023} but the validity of the KZM is not restricted to the classical dynamics. The Landau-Zener (LZ) transition formula \cite{landau_1932,zener_1932}, describing 
excitation probability in two-level systems, has been employed to analytically capture the KZM in the one dimensional 
driven transverse field quantum Ising model ($d=1, \nu=1, z=1$) \cite{dziarmaga_2005}.
KZM in quantum systems is studied in \cite{polkovnikov_2005,Uhlmann_2007,Uhlmann_2010,polkovnikov_2011,cucchietti_2007,dutta_2017,del_2018,sadhukhan_2020,schmitt_2022,dziarmaga_2022,dziarmaga_2023,kou_2023,naze_2023,huang_2024,yuan_2024} and there are experiments that have verified it for quantum phase transitions  \cite{chen_2011,cui_2016,cui_2020,bando_2020,higuera_2022,king_2022}.
Although KZM is believed to be broadly applicable, a conflicting observation has been reported in the study of ferroelectric 
phase transition where slower quenches lead to more topological defects when approaching the adiabatic limit \cite{griffin_scaling_2012}. 
This anti Kibble Zurek behavior (AKZ) can be the result of coupling to dissipative thermal bath \cite{nalbach_quantum_2015,patane_2008,rivas_2013} and also the difficulty of 
exactly controlling driven parameters \cite{ai_experimentally_2021,keesling_2018,braun_2015}. 

It has been shown that, adding the Gaussian white noise to the driven control parameter of a transverse field Ising model leads to the anti Kibble-Zurek behavior 
\cite{dutta_akzm}, which has also been observed in the transverse field $XY$ model \cite{gao_akzm,iwamura_2024}. The aforementioned theoretical prediction has been acknowledged 
experimentally by applying fully controlled noisy driving fields on the two level system with Landau-Zener crossings \cite{dziarmaga_2005,damski_2004,kayanuma_nonadiabatic,pokrovsky_fast_2003,kenmoe_2013} in a trapped ion simulator \cite{dutta_akzm,gao_akzm,ai_experimentally_2021}.

Despite numerous studies of defect generations in a wide variety of driven quantum systems in the presence of the uncorrelated (white) noise \cite{del_campo_2013,dutta_akzm,gao_akzm,ai_experimentally_2021,singh2021,Baghran2024}, comparatively little attention has been devoted to the driven quantum 
systems in the presence of the correlated (colored) noise \cite{Kiely2021}. From a physical point of view, the influence of noise correlation 
time on defect generation is an interesting open question: does a finite noise correlation time hinder or favor a noise-induced transition? 
In particular, we wish to determine how the dynamical features of the defect density is modified
according to the correlation of noise. In the presence of white noise, 
the scaling behavior of the defect density is known explicitly, which depends on the quench speed and the strength of noise \cite{dutta_akzm,gao_akzm}. We are going to investigate how the scaling behavior is changed whenever the noise is correlated in time.

To address the aforementioned questions we contribute to expand the systematic understanding of defect generation in driven transverse field $XY$ chain 
in the presence of Gaussian colored (Ornstein-Uhlenbeck) noise with perfect correlation in space, i.e noise fluctuation is the same for all spins at any time instant. The proposed conjecture, which is confirmed by our numerical simulations, provides a framework to distinguish the noise-induced excitations from the non-adiabatic excitation in AKZ mechanism.

It is worth to mention that in the pioneering works, Kayanuma \cite{kayanuma_nonadiabatic,kenmoe_2013} has addressed several applications, where the LZ theory was modified to take into account thermal noise 
or noise of a different nature.
Such results depend on whether the noise is slow or fast. More precisely, it depends on the noise correlation 
time whether it is longer or shorter than the Landau-Zener transition time, $\tau_{LZ}$. 
Nevertheless, to the best of our knowledge, there is no comprehensive study to manifest the effects of noise correlation time in KZM.

The outline of this work is as follows. In section \ref{sec:level1}, we briefly review the exact solution and phase diagram 
of the transverse field $XY$ model.  To calculate the transition probabilities in the presence of colored noise, the exact master 
equation of a quantum system perturbed by colored noise is presented in Section \ref{sec:level2}.
Sections \ref{sec:level3} and \ref{sec:level4}, are dedicated to the numerical simulations of the model based on the exact master
equation, where we present our conjecture which captures the noise-induced excitation features.
Finally, Section \ref{conclusion} contains some concluding remarks.


\section{Transverse field XY Model}\label{sec:level1}
In this section we review the phase diagram of the exactly solvable transverse field XY Model, which has the following Hamiltonian
%
\begin{equation}
H = - J \Big( \sum_{j}^{N} \big( \frac{1+\gamma(t)}{2} \sigma_{j}^{x} \sigma_{j+1}^{x} 
+ \frac{1-\gamma(t)}{2} \sigma_{j}^{y} \sigma_{j+1}^{y} \big) +g(t) \sum_{j}^{N} \sigma_{j}^{z} \Big),
\end{equation}
%
where $\sigma_{j}^{x,y,z}$ are Pauli matrices, and $\gamma$ is the anisotropy coefficient which describes 
the difference of interactive strength in the $x$ and $y$ components. 
We consider periodic boundary condition and without loss of generality we set $J=1$ as the energy
scale and the system size is $N=1000$.

The system reduces to the transverse field Ising model and isotropic $XY$ chain for $\gamma=1$ and $\gamma=0$, respectively.  
This Hamiltonian can be mapped to the spinless free fermions by using the Jordan-Wigner transformation \cite{lieb_two_1961,franchini_introduction_2016} ,
%
\begin{eqnarray}
\label{free_fermion}
H &=&- \sum_{j}^{N} \Big( \gamma(t) [c_{j}^{\dagger} c_{j+1}^{\dagger} - c_j c_{j+1}] + c_{j}^{\dagger} c_{j+1} - c_j c_{j+1}^{\dagger} \Big)
\nonumber \\
& & - g(t) \sum_{j}^{N} ( 2 c_{j}^{\dagger} c_j - 1) .
\end{eqnarray}
%
In the momentum space, by introducing the Nambu spinor $\Psi_{k}^{\dagger} \equiv (c_{k}^{\dagger} \; c_{-k})$ within the even parity subspace,
the free fermion Hamiltonian Eq. (\ref{free_fermion}) can be written as the sum of $N$ non-interacting
terms $H = \sum_{k>0} \Psi_{k}^{\dagger} \mathcal{H}_k(t) \Psi_{k}$, where $k =(2m-1)\pi/N$ with $m=1, 2, \ldots, N/2$ and the local Hamiltonian reads:
%
\begin{equation}
\mathcal{H}_k(t) = 2 \big( g(t) - \cos(k) \big) \hat{\sigma}^{z} + 2\gamma(t) sin(k) \hat{\sigma}^{x}.
\label{Hk}
\end{equation}
%
%
\begin{figure}
\centerline{\includegraphics[width=\columnwidth]{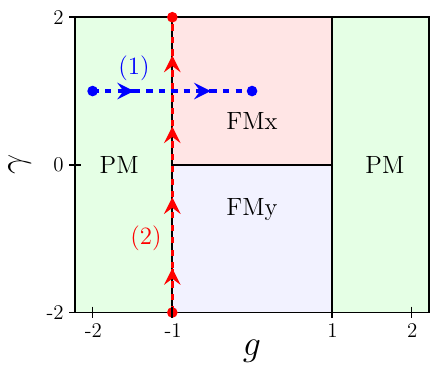}}
\centering
\caption{(Color online) Quantum phase diagram of the spin-$1/2$ $XY$ chain, where $g$ is the 
transverse field and  $\gamma$ is the anisotropy coefficient. 
The horizontal blue dashed-line (1) represents the path of the transverse filed quench and 
the vertical red dashed-line (2) shows the path of anisotropy quench.}
\label{phase}
\end{figure}
%

The phase diagram of time-independent $XY$ model has been depicted in Fig. \ref{phase}. The model reveals two critical points 
at $g_c=\pm1$, which separates the paramagnetic phase from the ferromagnetic ones. In addition the line $\gamma=0$ for $-1<g<1$ 
corresponds to the anisotropic critical line separating the two ferromagnetic phases FMx and FMy, with ordering in $x$ and $y$
directions, respectively. Moreover, there are two gapless lines along $g=\pm1$ where a multicritical point located at $g=\pm1, \gamma=0$.
A quantum phase transition occurs when the system is driven by one or more parameters across the transition lines/points on the phase diagram.

For the convenience, and without loss of generality, we choose only one of the Hamiltonian's parameters ($g, \gamma$) as linearly quenched in time and the rest is fixed during the quench protocol. 

In the following sections we will investigate two different quench protocols between different regions on the phase diagram in the presence of a Gaussian colored noise \cite{Jafari2024}.
First, we consider the transverse field quench, where only $g(t)$ is varied in time, which is quenching across the boundary line $g_c=-1$ between
the paramagnetic and ferromagnetic phases, which is denoted by path-(1) in Fig. \ref{phase}. Then we study the quench along the gapless line ($g=-1$), only $\gamma(t)$ is varied and crosses the multicritical point ($g=-1,\gamma=0$), which is represented by path-(2) in Fig. \ref{phase}.


\section{Ensemble-averaged transition probabilities: Exact noise master equation}\label{sec:level2}

For transparency and ease of notation, we considered a general time-dependent Hamiltonian
%
\begin{equation}
H = H_0(t)  + \eta (t) H_1(t),
\label{H0H1}
\end{equation}
%
where $H_{0}(t)$ is the noise-free Hamiltonian, while $H_{1}(t)$ is the ``noisy" part of Hamiltonian where  $\eta(t)$ 
is a Gaussian colored noise with zero mean and the following temporal correlation 
%
\begin{figure*}		
\centerline{\includegraphics[width=0.5\linewidth]{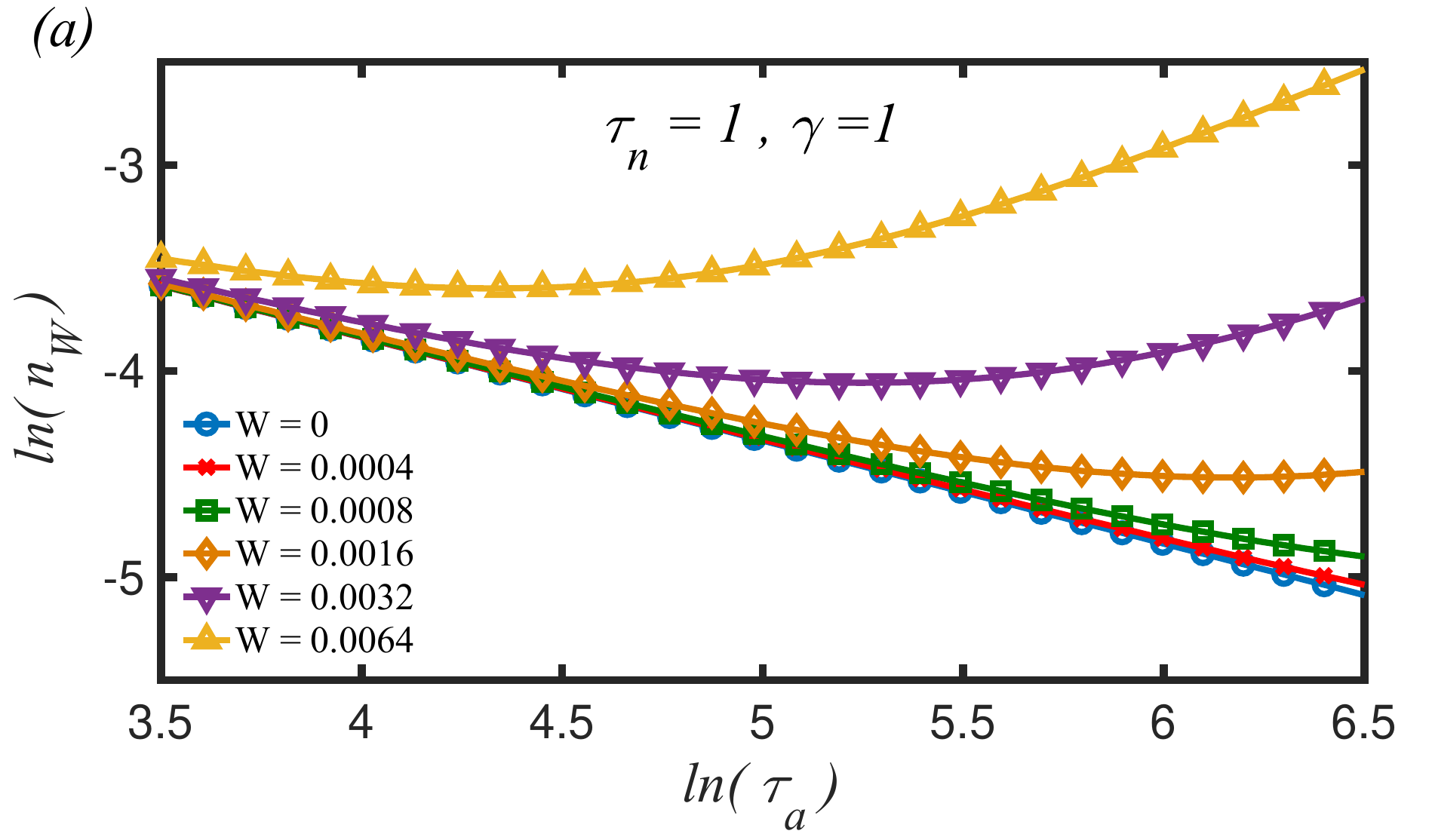}
\includegraphics[width=0.5\linewidth]{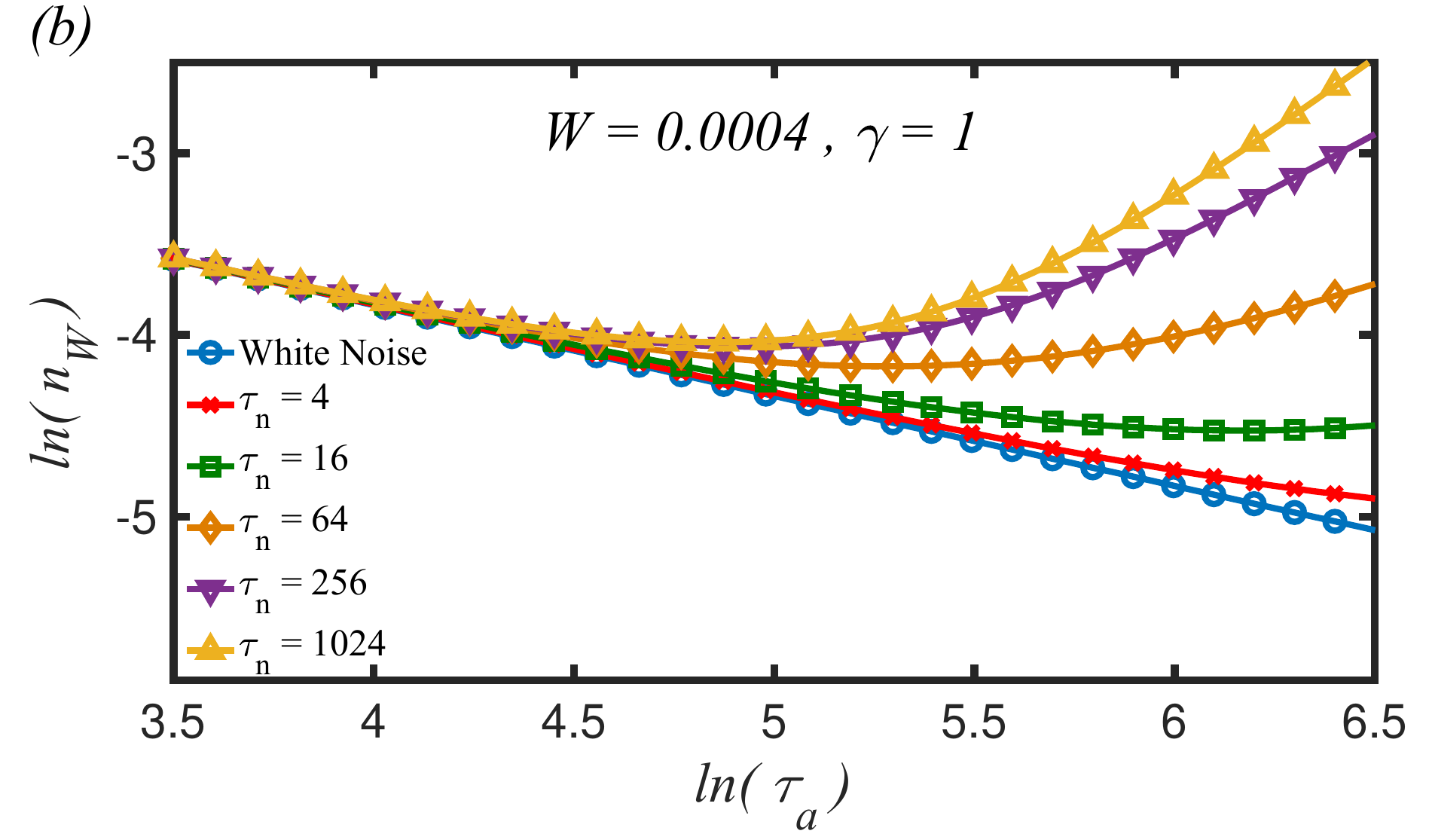}}
\centering
\caption{(Color online) The density of excitations ($n_W$) vs. annealing time ($\tau_a$) of the
transverse field Ising model ($\gamma=1$) for (a) different noise power ($W$) and $\tau_n = 1$,
and (b) $W = 0.0004$ and different values of noise correlation time.}
\label{nex_ising}
\end{figure*}
%
%
\begin{equation}
R(\tau) = \langle \eta(t) \eta(t+\tau) \rangle = \frac{\xi^{2}}{2 \tau_{n}}\exp(-|\tau| / \tau_{n}) ,
\label{noise_corr}
\end{equation}
%
where $\tau_{n}$ is the noise correlation time, $\xi$ the noise amplitude and the total power of noise is given by $W^{2} = \xi^2/2 \tau_{n}$.

As shown in Refs.\cite{Jafari2024,Kiely2021,budini_2001} the noise-averaged density
matrix $\rho(t)= \langle \rho_{\eta}(t) \rangle$ over the whole noise distribution, is given by the following (nonperturbative) exact master equation  
%
\begin{eqnarray}
	&&\frac{d}{dt} \rho (t) =  -i [ H_0(t) , \rho (t)] \nonumber \\
	&&   - \frac{\xi^{2}}{2 \tau_{n}} \Big[H_{1, k}(t) , \int_{t_i}^{t} e^{- \frac{|t-s|}  { \tau_{n}}}   [H_{1, k} (s) , \rho (s) ] ds \Big].
	\label{master}
\end{eqnarray}
%

As mentioned, the one-dimensional $XY$ model can be written as a sum of decoupled $k$-mode Hamiltonians \cite{Barouch_1970,Jafari2024,nalbach_quantum_2015} and consequently the density matrix of the model 
has a direct product structure $\rho(t) = \otimes_{k} \rho_{k}(t)$. Therefore, the noise master equation for the ensemble-averaged
density matrix $\rho_{k}(t)$ takes the form
%
\begin{eqnarray}
&&\frac{d}{dt}  \rho_k (t) =  -i [ H_{0,k}(t) , \rho_k (t)] \nonumber \\
  &&- \frac{\xi^{2}}{2 \tau_{n}} \Big[H_{1, k}(t) , \int_{t_i}^{t} e^{- \frac{|t-s|}  { \tau_{n}}}   [H_{1, k} (s) , \rho_k (s) ] ds\Big] .
\label{master2}
\end{eqnarray}
%

It should be mentioned that, the above master equation reduces to the white noise master equation \cite{Jafari2024,dutta_akzm,Kiely2021}
%
\begin{eqnarray}
\no
	\frac{d}{dt}  \rho_k (t)=  -i [ H_{0,k}(t) , \rho_k (t)] -  \frac{\xi^{2}}{2 \tau_{n}} \Big[H_1(t) ,   [H_1(t), \rho_k (t) ]\Big],
\end{eqnarray}
%
by taking the limit of $\tau_n \rightarrow 0$.

By translating Eq. (\ref{master2}) into two coupled differential equations, the mean transition probabilities are obtained numerically 
as ensemble averages $\langle p_k\rangle$ over the noise distribution $\{\eta\}$ \cite{Jafari2024} (for details see Appendix \ref{APA}). 


\section{Linear time dependent transverse field}\label{sec:level3}

Here, we fix the anisotropy ($\gamma(t)=\gamma$) to search for the defect density 
generation due to the sweep of noisy time dependent transverse field i.e., $g(t) = g_0 (t) + \eta (t)$ through 
the critical point $g_c=-1$. The first term in the noisy transverse field shows the linear time dependent 
deterministic part $g_0 (t) = g_i + (g_f - g_i)t/\tau_a$ which changes from $g_i$ at the initial time $t_i =0$ to the 
final value $g_f$ at $t_f =\tau_a$, where $\tau_a$ is 
the annealing time. We should mention that the quench time scale defined as the inverse quench rate $\tau_{Q} = 1/(dg_0 (t)/dt) = \tau_a / (g_f - g_i)$, 
which reveals that the scaling behavior for both $\tau_a$ and $\tau_Q$ is the same.
The second term $\eta (t)$ represents the stochastic Gaussian part of the control field with zero mean $\langle \eta (t) \rangle = 0$ 
and canonical Ornstein-Uhlenbeck \cite{cox2017} two-point correlations  given by Eq. (\ref{noise_corr}). 
We should note that, in the mentioned ramp quench protocol, the Landau-Zener transition time is proportional to the 
annealing time, i.e., $\tau_{LZ}\propto\tau_a$. Accordingly, fast noise corresponds to $\tau_{n}\ll\tau_a$ and in the opposite 
limit the slow noise is represented by $\tau_{n}\gg\tau_a$.\\

To apply the exact noise master equation (Eq. (\ref{master2})) we should decouple the Hamiltonian into noiseless and 
noisy parts $H = H_0 (t) + \eta (t) H_1$.
Using Eq \eqref{Hk} and definition $g(t) = g_0 (t) + \eta (t)$, the noiseless and noisy parts of the full Hamiltonian are given as $H_0 = \sum_{k>0} \Psi_{k}^{\dagger} \mathcal{H}_{0,k}(t) \Psi_{k}$, where 
$\mathcal{H}_{0,k}(t) = 2(g_0(t) - \cos(k) ) \hat{\sigma}^{z} +  2 \gamma sin(k) \hat{\sigma}^{x}$ and 
$H_1 = \sum_{k>0} \Psi_{k}^{\dagger} \mathcal{H}_{1,k} \Psi_{k}$, with $\mathcal{H}_{1,k}=2 \hat{\sigma}^{z}$.
By numerically solving the master equation and acquiring the average density matrix $\rho_{k}(\tau_a)$, the density 
of excitations is given by the expectation values of the noise-averaged density matrix ($\rho_{k}$), i.e.,
%
\begin{equation}
n_W =  \frac{2}{N} \sum_{k>0} \langle \varepsilon_{k}(t_f)| \rho_{k}(t_f) |\varepsilon_{k}(t_f)\rangle,
\label{nw}
\end{equation}
%
where $|\varepsilon_k(t_f)\rangle$ is the excited eigenstate of Hamiltonian at the end of quench, $t=t_f$.

%
\begin{figure*}		
\centerline{\includegraphics[width=0.5\linewidth]{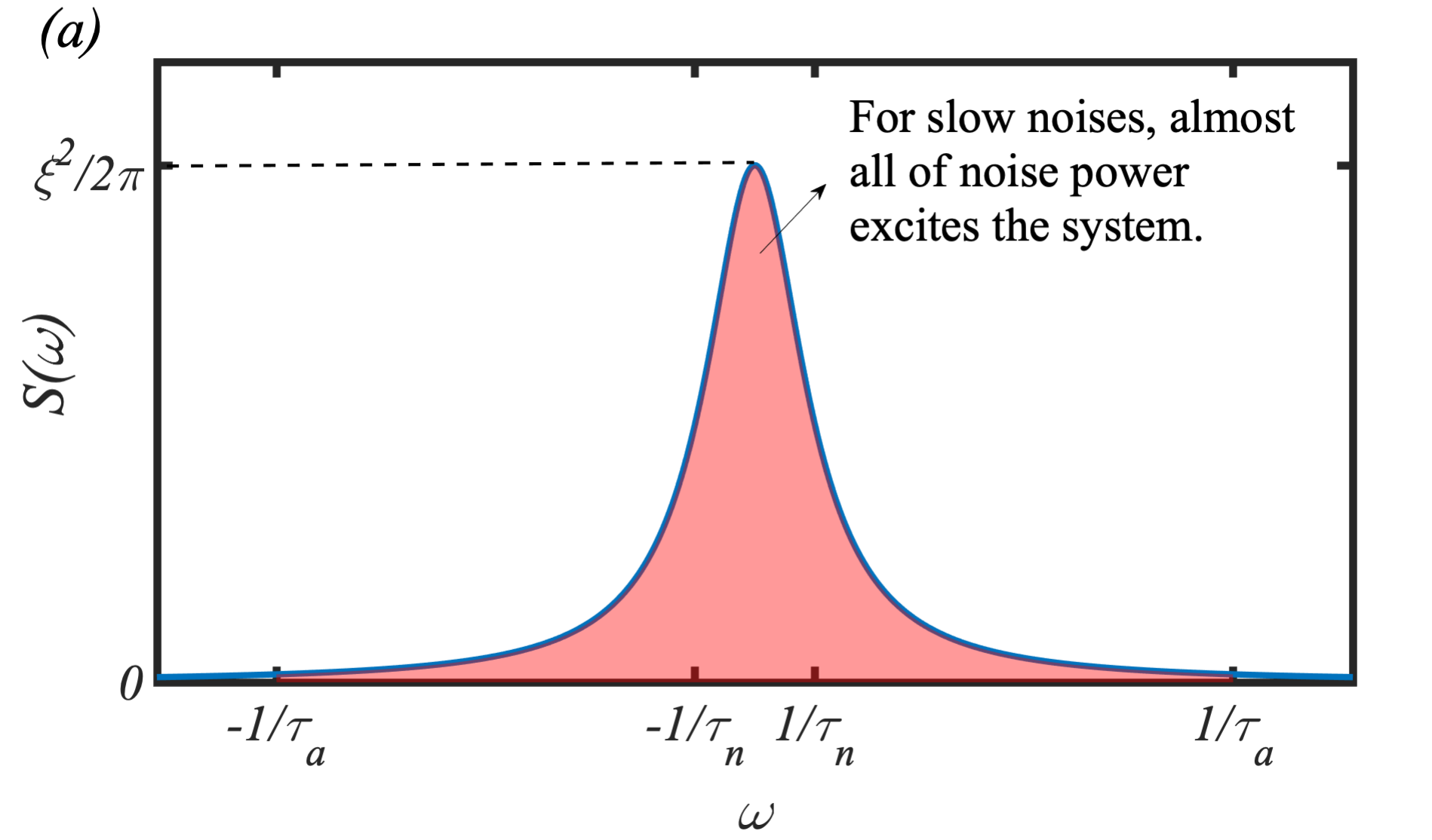}
\includegraphics[width=0.5\linewidth]{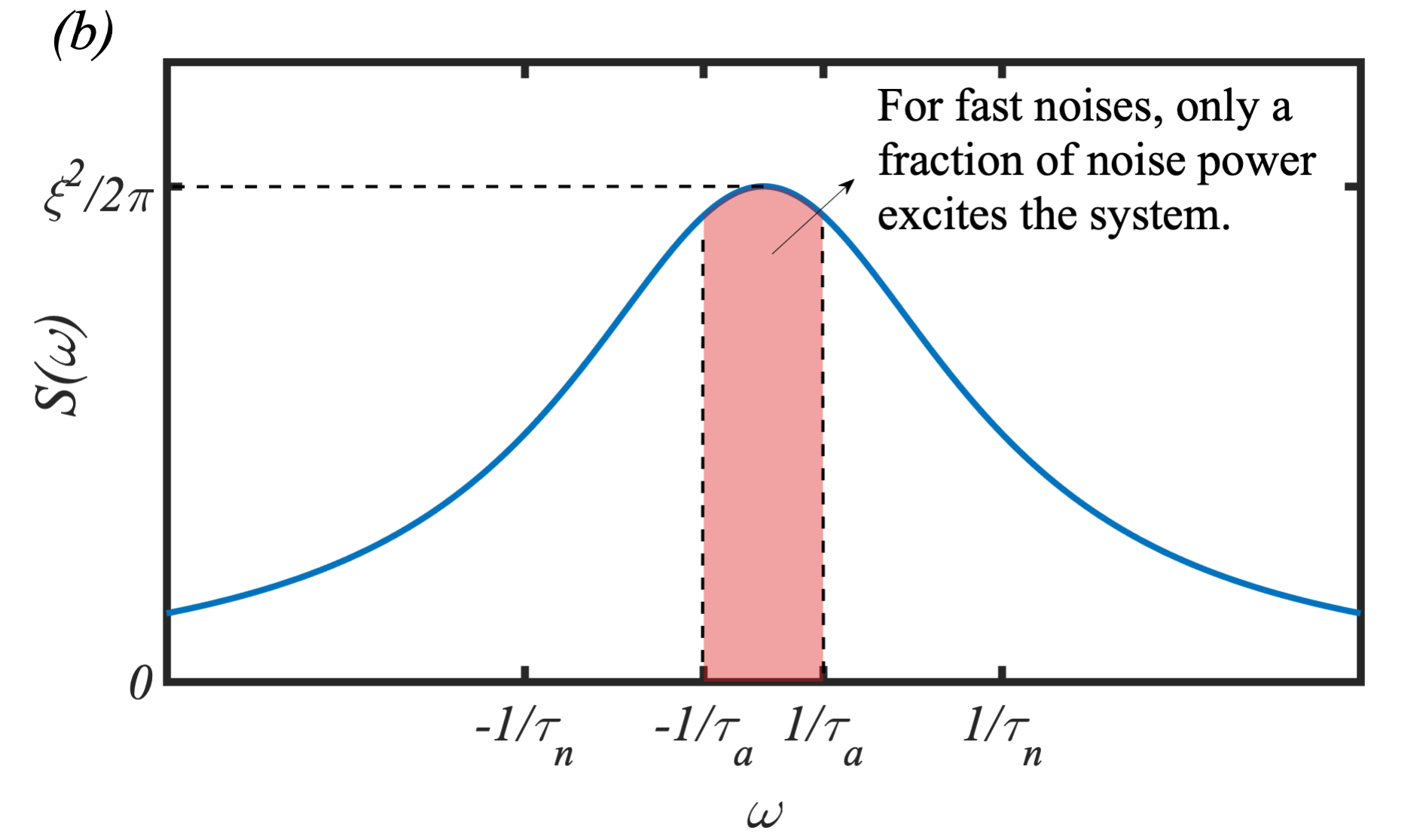}}
\centering
\caption{(Color online) The spectrum of noise power for (a) slow noises ($\tau_n \gg \tau_a$), where almost all of spectral components excites the system and (b) fast noises ($\tau_n \ll \tau_a$), where only part of noise spectral components excites the system.}
\label{PowerSpectrum}
\end{figure*}
%
\subsection{Crossing the critical point $g_c=-1$ for $\gamma = 1$} \label{sec:level3a}

As mentioned, the transverse field $XY$ model reduces to the transverse field Ising chain, for $\gamma=1$. 
Here we assume that $g(t)$ changes linearly from $g_i = -2$ to $g_f = 0$, where the system crosses the critical 
point $g_c=-1$ (blue dashed-line in Fig. \ref{phase}) with annealing time $\tau_a$. 
The excited eigenstate of the model at the end of the quench ($g_f = 0 $), is given by
%
\begin{equation}
	|\varepsilon_{k}(t_f)\rangle = 
	\begin{pmatrix}
		sin(k/2) \\
		cos(k/2)
	\end{pmatrix}.
\end{equation}
%

The results of numerical simulations based on exact master equation have been plotted in Fig. \ref{nex_ising}. The numerical results 
show good agreement with KZM theoretical prediction in the noiseless case $W=0$ (Fig. \ref{nex_ising}(a)), where the density of excitations $n_W$ scales as a power of the annealing time with the exponent $\delta=-1/2$, i.e., $n_W \propto \tau_{a}^{-1/2}$.

In Fig. \ref{nex_ising}(a) the density of defects has been plotted versus the annealing time $\tau_a$  
for small noise correlation time $\tau_n=1$ (approximately equivalent to the white noise) and different values of the noise power $W^2$. 
As we observe, the density of excitations grows by increasing the annealing time in the presence of noise, 
while according to the KZM, we expect that the system gets less excited for larger $\tau_a$ (adiabaticity). 
This unexpected behavior originates from the noise effects and can be interpreted as the noise-induced excitations. (for details see Appendix \ref{APB}.)

In other words, the defect generation is controlled by two dissenting mechanisms: (i) noise-induced excitations, which enhances by increasing $\tau_a$ 
and (ii) non-adiabatic excitations, which is suppressed by raising the annealing time and quench time scale. We expect that the non-adiabaticity by small values of the quench time scale 
gives less time for noise to become effective. Consequently, the interplay of these two competing effects results in a minimum of the density of defects 
at optimum annealing time $\tau^{(opt)}_a$.

Fig. \ref{nex_ising}(b) displays the density of excitation for different values of the noise correlation time for a fixed noise power ($W=0.0004$). 
As seen, for constant noise power, increasing the noise correlation time leads to more excitations. This means that large noise correlation time 
($\tau_n \gg \tau_a$) results in the accumulation of noise power at low frequency range, i.e frequencies below $1/\tau_a$ (Fig. \ref{PowerSpectrum}(a)). 
However, in the fast noise case, i.e., small noise correlation time ($\tau_n \ll \tau_a$), the whole noise power is not effective
in the excitation process and only a fraction of noise power takes part in defects formation (Fig. \ref{PowerSpectrum}(b)). 

\begin{figure*}[!t]
\centerline{\includegraphics[width=0.5\linewidth]{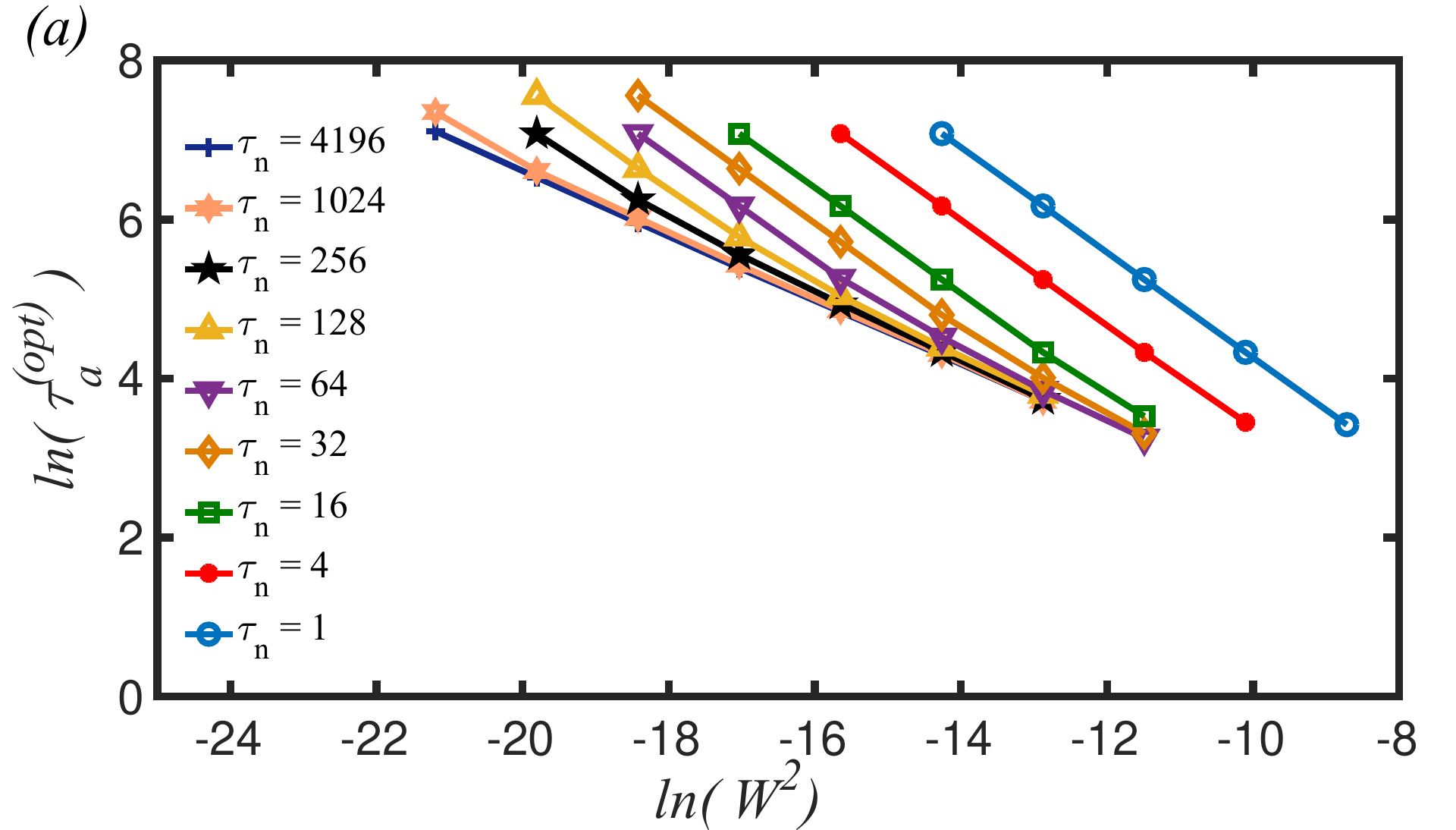}
\includegraphics[width=0.5\linewidth]{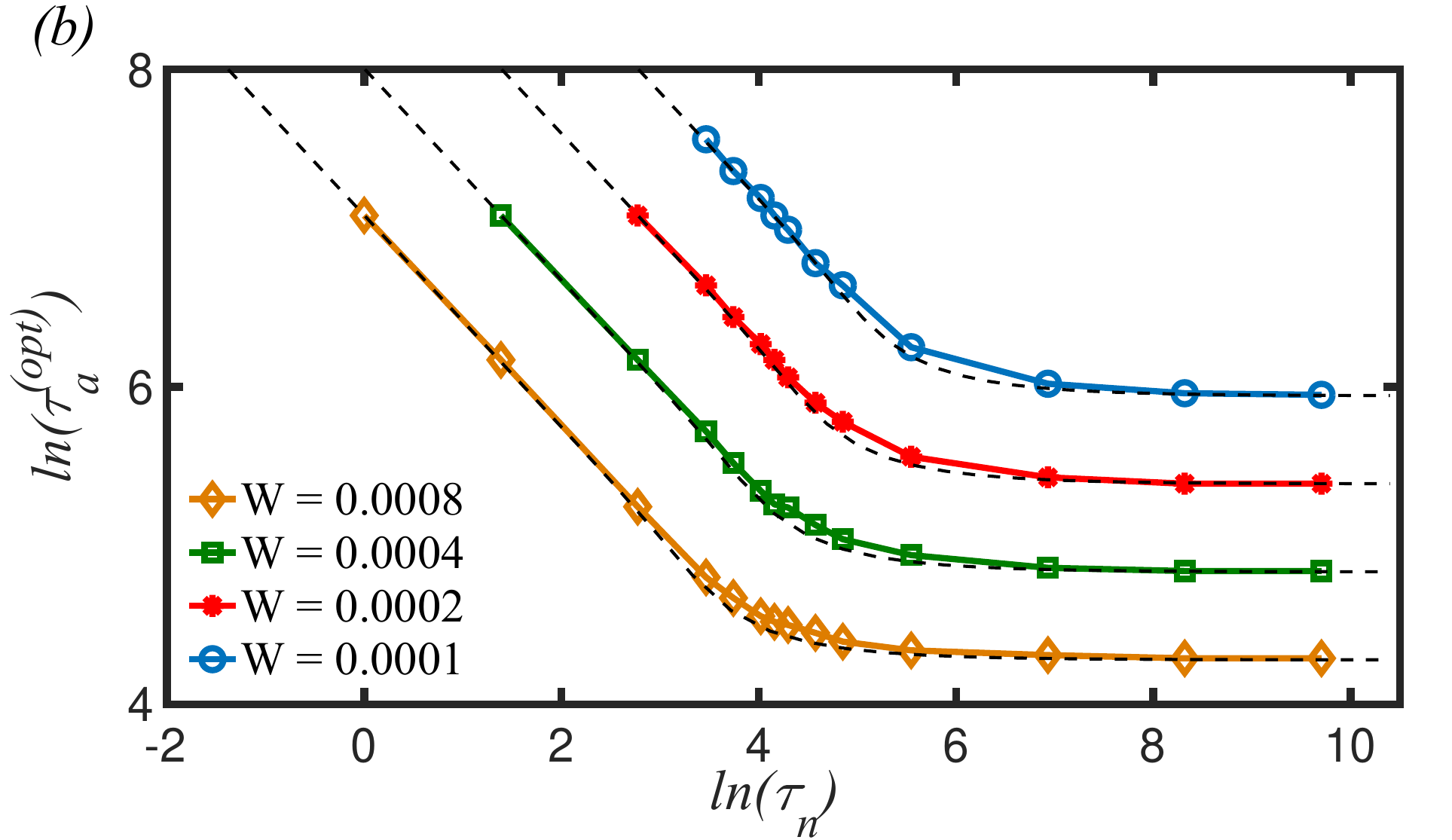}
}
\centering
\caption{(Color online) The scaling of $\tau_{a}^{(opt)}$ versus (a) noise power and 
(b) noise correlation time, both for $\gamma = 1$. The dashed lines display the analytic 
conjecture (Eq. \eqref{generaltauopt}) with the coefficients $a=1/2\pi$, $b = 0.88$ and $c = 3.40$ and the solid lines illustrate our numerical simulations.
}
\label{scaling}
\end{figure*}

The density of excitations as a function of annealing time has a minimum (as seen in Fig. \ref{nex_ising}), which occurs at the optimal annealing time ($\tau^{(opt)}_{a}$).
The scaling of optimal annealing time versus the noise power ($W^2$) has been plotted in Fig. \ref{scaling}(a) for different values of the noise correlation time. 
The numerical results show that, for fast noise, $\tau^{(opt)}_a$ indicates a power law scaling with noise power $W^2$, i.e., 
$\tau^{(opt)}_a \propto (W^2)^{\beta}$, where $\beta\approx -0.66$, which is the same as the corresponding one in the presence of white noises \cite{dutta_akzm,gao_akzm}. 
The scaling exponent of optimal annealing time is different for slow noise, which is derived to be $\beta\approx -0.4$.

Moreover, we have investigated the scaling behavior of optimal annealing time in terms of the noise correlation time. 
Fig. \ref{scaling}(b) represents the scaling of the optimal annealing time versus the noise correlation time for different values of noise power. 
For small noise correlation time, all curves show power law scaling of $\tau_n$ with the same exponent $\beta\approx-0.66$.
However, for very large noise correlation time, the optimum annealing time makes a plateau versus $\tau_n$ which means that the optimal annealing time of the slow noises is independent of the noise correlation time.

%
Having identified the effect of noise correlation time on the defect density, we propose the analytical conjecture to 
better understand the noise induced-excitation features and scaling. To this end, we implement the power spectrum of noise, which 
is equal to Fourier transform of the noise correlation \cite{uhlenbeck_1945,mallick_2007}
%
\begin{equation}
S(\omega) = \mathcal{F} \lbrace R(\tau) \rbrace = \dfrac{\xi^2}{2 \pi (\omega^2 \tau_{n}^{2} + 1)}.
\end{equation}
%
The spectral components of noise with frequency less than quench characteristic frequency $\omega_{c} = c/\tau_a$ (Landau-Zener time scale), 
where $c$ is a constant \cite{pokrovsky_fast_2003}, contribute in the excitation formation. The excitation rate which is proportional 
to the average of noise power within the frequency interval below $\omega_c$ is given by
%
\begin{equation}
S_{av} = \frac{\int_{-\omega_c}^{\omega_c} S(\omega) d\omega}{2 \omega_c} = \frac{ W^2 \tau_a}{\pi c } \arctan(\frac{c \tau_n}{\tau_a}).
\end{equation}
%
Therefore, the noise-induced excitations ($n_{noise}$) in the quench interval $t \in\, [t_i=0,t_f=\tau_a]$, which is proportional to $S_{av} \tau_a$,
is obtained to be
%
\begin{equation}
n_{noise}  = b W^2 \tau_{a}^{2}\arctan(\frac{c \tau_n}{\tau_a}),
\label{noise1}
\end{equation}
%
where $b$ is a constant. The total density of excitations is the sum of noise-induced excitations and excitations produced due to KZM ($n_{0}$), i.e.,
%
\bea
\no
n_W = n_{0} +n _{noise} = a \tau_{a}^{-1/2} + b W^2 \tau_{a}^{2} \arctan (\dfrac{c \tau_n}{\tau_a}).\\
\label{generalnoise}
\eea
%
As mentioned earlier, the defect density reaches a minimum at the optimum annealing time. Thus, the optimum annealing time is determined by
the roots of equation $dn_W/d\tau_a=0$, i.e.,
%
\begin{equation}
\frac{1}{2}a \tau_{a}^{-3/2} + bcW^2 \tau_n ( \frac{1}{1 + (c \tau_n/\tau_a)^2} - \frac{2 \arctan(c \tau_n/\tau_a)}{c \tau_n / \tau_a}) = 0.
\label{generaltauopt}
\end{equation}
%

It is worthy to mention that the coefficient $a$ can be calculated analytically using LZ formula for noiseless case ($W=0$), where $p_k$ is given by
%
\bea
\label{LZ1}
p_k = e^{-2 \pi  \tau_Q k^2},
\eea
%
which results in
%
\bea
\label{LZ2}
n_0 =\frac{1}{\pi} \int_{0}^{\pi} p_k dk = \frac{\tau_{Q}^{-1/2}}{2 \pi \sqrt{2}}.
\eea
%
By using the quench time scale definition $\tau_Q = \tau_a/(g_f-g_i)= \tau_a / 2$, and comparing Eq. (\ref{LZ2}) with Eq. (\ref{generalnoise}) 
the mentioned coefficient is obtained to be
\bea
a=\frac{1}{2 \pi} . 
\eea
The other coefficients $b$ and $c$ are calculated based on our numerical results which are $b=0.88$ and $c=3.40$.

The dashed black lines in Fig. \ref{scaling}(b) represent the optimum annealing time calculated based on the analytical 
conjecture Eq. (\ref{generaltauopt}) with $a = 1/2\pi$, $b = 0.88$ and $c = 3.40$, which is in good agreement with the numerical results.
We have also compared the density of excitation obtained by Eq. \eqref{generalnoise} with numerical results in Appendix \ref{APC}.
 
To better understand the analytical conjecture and effects of noise-induced excitations in defect density, 
we examine Eq. \eqref{generaltauopt} for two extreme cases: (I) fast noise, and (II) slow noise.

\begin{enumerate}[label=(\Roman*)]

\item  \textbf{Fast noise ($\tau_n \ll \tau_a$)}:
In this limit, the $\arctan(c \tau_n/\tau_a)$ function can be approximated by $c \tau_n/\tau_a$, and Eq. \eqref{noise1} is reduced
to 
%
\bea
\label{eq15}
n _{noise} \overset{\tau_n \ll \tau_a}{\approx} b c W^2 \tau_{a} \tau_n,
\eea
%

It is obvious that for fast noises, the contribution of noise-induced excitation in defect generation ($n_{noise}$) scales linearly with the noise correlation time $\tau_n$ 
and annealing time $\tau_a$ and also with the noise power ($W^2$). 
The noise-induced excitations, obtained by numerically solving the exact master equation, has been plotted versus annealing time in Fig. \ref{fast}(a) for different values of noise power and $\tau_n =1$. The numerical results show that the 
slope of lines is proportional to noise power $W^2$. Moreover, Fig. \ref{fast}(b) represents the noise-induced excitations versus 
annealing time for different values of fast noise correlation time (i.e. $\tau_n =1, 4, 16$), for $W=0.0002$, where the slope of lines 
vary linearly with the noise correlation time $\tau_n$. As clear, the numerical simulations is consistent with our proposed conjecture.
  
%
\begin{figure*}
\centerline{\includegraphics[width=0.5\linewidth]{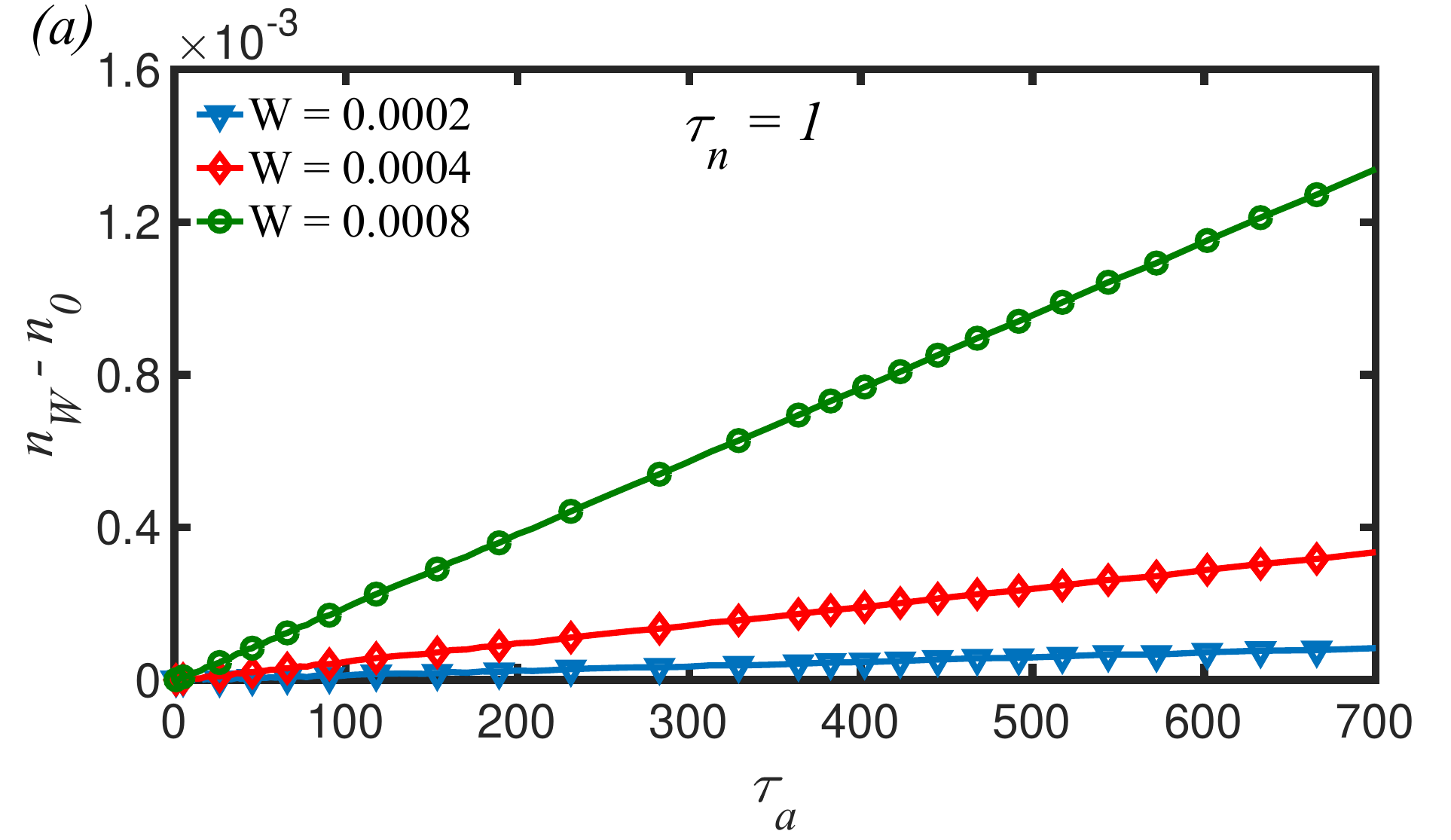}
\includegraphics[width=0.5\linewidth]{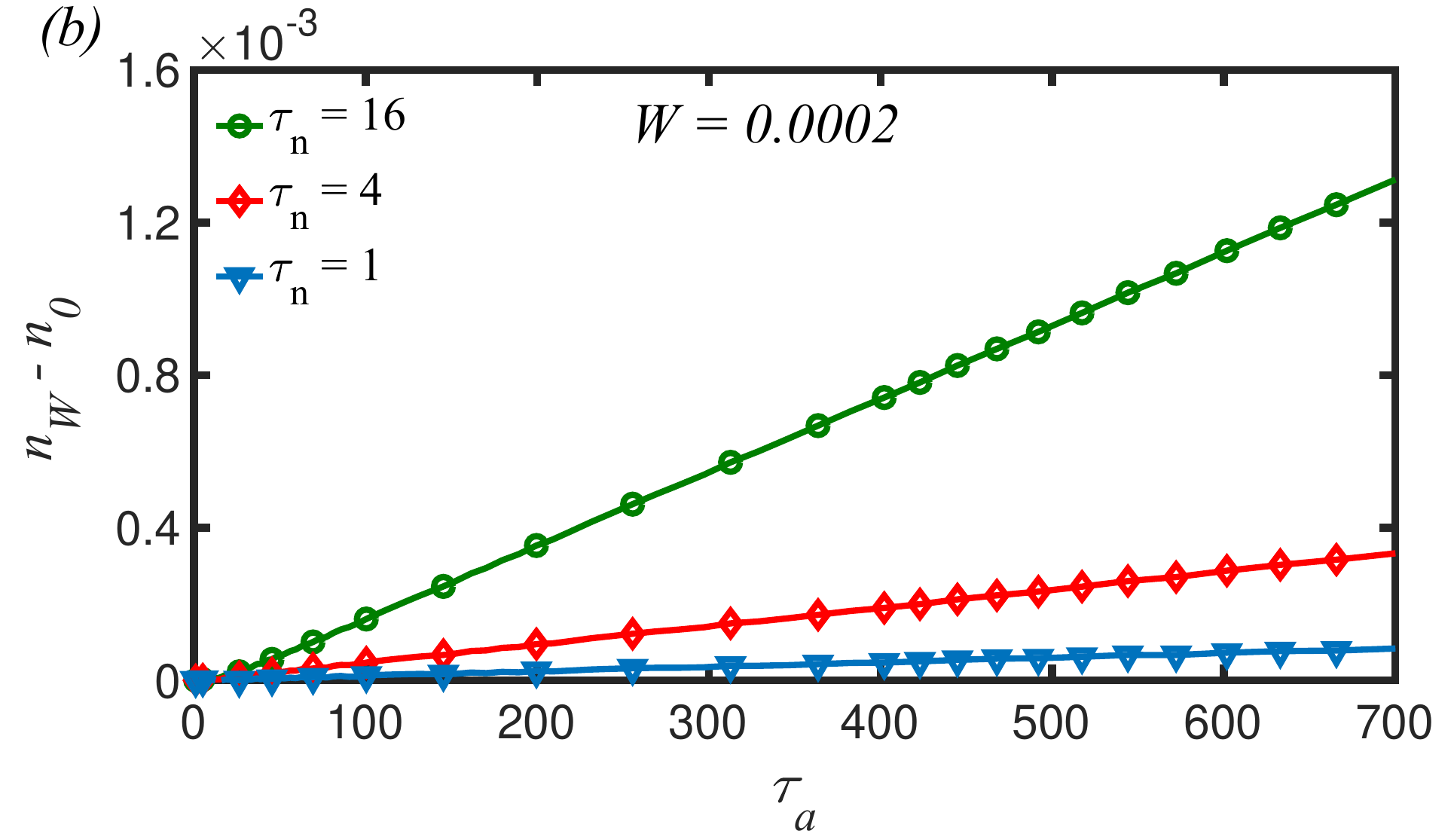}
}
\centering 
\caption{(Color online) Fast noise case: The noise-induced contribution to excitations ($n_W-n_0$)
vs. $\tau_a$ for (a) constant noise correlation time and different noise power and (b) constant power and different noise correlation time.
The noise-induced density changes linearly with $\tau_a$, $\tau_n$ and $W^2$, because at low frequencies,
the spectrum of noise is almost constant.}
\label{fast}
\end{figure*}
%

%
\begin{figure*}
\centerline{
\includegraphics[width=0.5\linewidth]{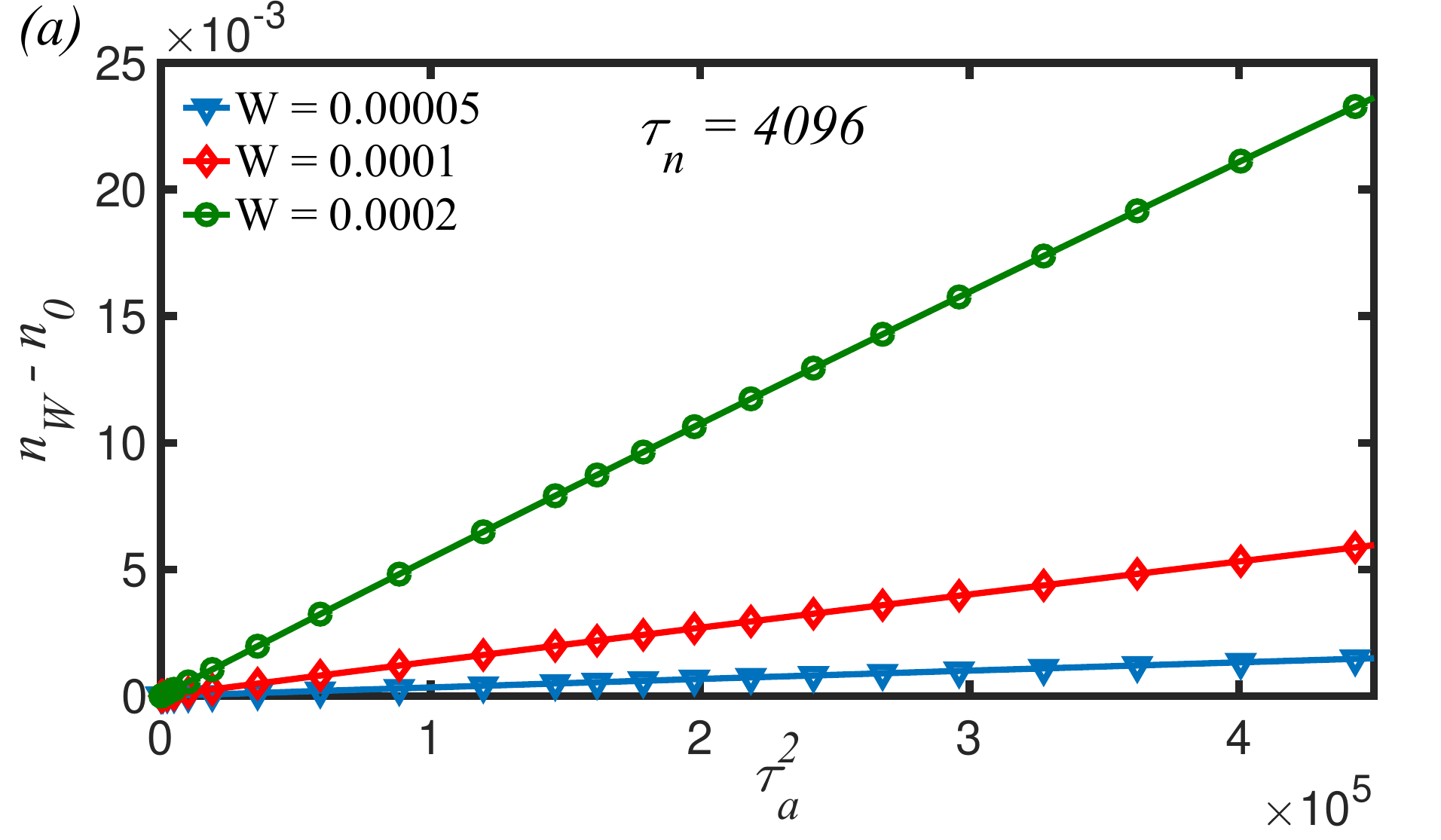}
\includegraphics[width=0.5\linewidth]{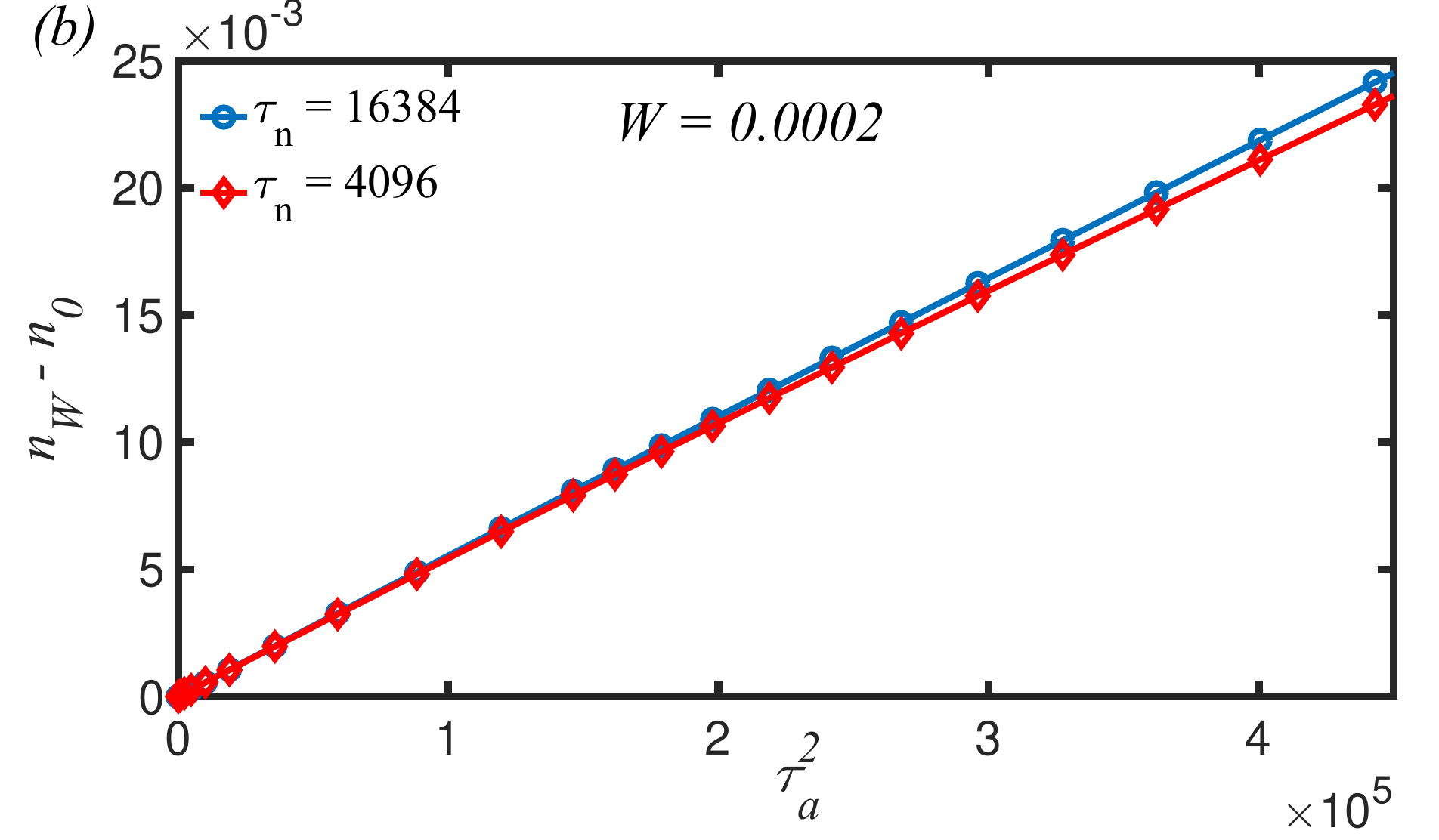}
}
\centering  
\caption{(Color online) Slow noise case: The noise-induced contribution to excitations
vs. $\tau_{a}^{2}$, for (a) constant noise correlation time and different power of noise and (b) constant power and different noise correlation time.
The noise-induced density
changes linearly with  $W^2$ and depends on the annealing time quadratically (i.e. linear in 
$\tau_{a}^{2}$). Here, all of noise power excites the system.}
\label{slow}
\end{figure*}
%
Furthermore, the optimal annealing time for fast noise limit is obtained by solving
$\frac{d n_W}{d \tau_a} = 0$, which results in
%
\bea
\label{eq16}
\tau_{a}^{(opt)} \propto (W^2 \tau_n)^{-2/3} \propto (\xi^2)^{-2/3}.
\eea
%
The above equation suggests that the scaling exponent of optimum annealing time is $\beta=-2/3$, which is consistent with 
the numerical results for fast noises $\beta\approx-0.66$ as given in Fig. \ref{scaling}.

\item  \textbf{Slow noise ($\tau_n \gg \tau_a$)} :
In the slow noise limit , the $\arctan(c\tau_n/\tau_a)$ is approximately equal to $\pi/2$, where
the density of defects (Eq. \eqref{noise1}) is reduced to 
%
\bea
\label{eq17}
n _{noise} \overset{\tau_n \gg \tau_a}{\approx} \frac{b \pi}{2} W^2 \tau_{a}^{2} .
\eea
%
It is clearly seen that, in contrast to the fast noise case, the density of defects does not depend on the noise correlation time for slow noises. 
In addition, for slow noises, the noise-induced excitation scales linearly with the square of annealing time
($\tau^2_a$), while in the presence of the fast noises it follows linear scaling with $\tau_a$. However, scaling of the noise-induced 
excitations with noise power is the same for both slow and fast noises. The noise contribution to the defects density has been shown 
in Fig. \ref{slow}(a) versus $\tau^2_a$, for different values of noise power, for slow noise $\tau_n =4096$. As seen, the slow noise 
contribution in defects density scales linearly with $\tau^2_a$. 
More detailed analysis uncovers that the slope of lines in Fig. \ref{slow}(a) is proportional to $W^2$. Fig. \ref{slow}(b) also shows that 
the slow noise contribution to the excitations for noise power $W=0.0002$ is approximately the same for all noise correlation times $\tau_n$.

For slow noises, the scaling of optimal annealing time with the noise power $W$, based on our conjecture, is obtained as follows
%
\begin{equation}
\label{eq18}
\frac{d n_W}{d \tau_a} = 0 \Rightarrow \tau_{a}^{(opt)} \propto (W^2)^{-2/5},
\end{equation}
%
which yields $\beta=-2/5$. Moreover, the optimum annealing time is independent of the noise correlation time. These results 
show good agreement with the numerical simulations shown in Fig. \ref{scaling}.

\end{enumerate}

\subsection{Crossing the critical point $g_c=-1$ for $\gamma \neq 1$}
To investigate the effect of anisotropy on the noise-induced transitions, here, we have considered 
once again $g(t)$ changes linearly from $g_i = -2$ at $t_i=0$ to $g_f = 0$ at $t_f=\tau_a$, where the system crosses the critical 
point $g_c=-1$ for $\gamma \neq 1$. At the end of the quench, the non-normalized excited state of the model is described as

%
\begin{figure*}
\centerline{\includegraphics[width=0.5\linewidth]{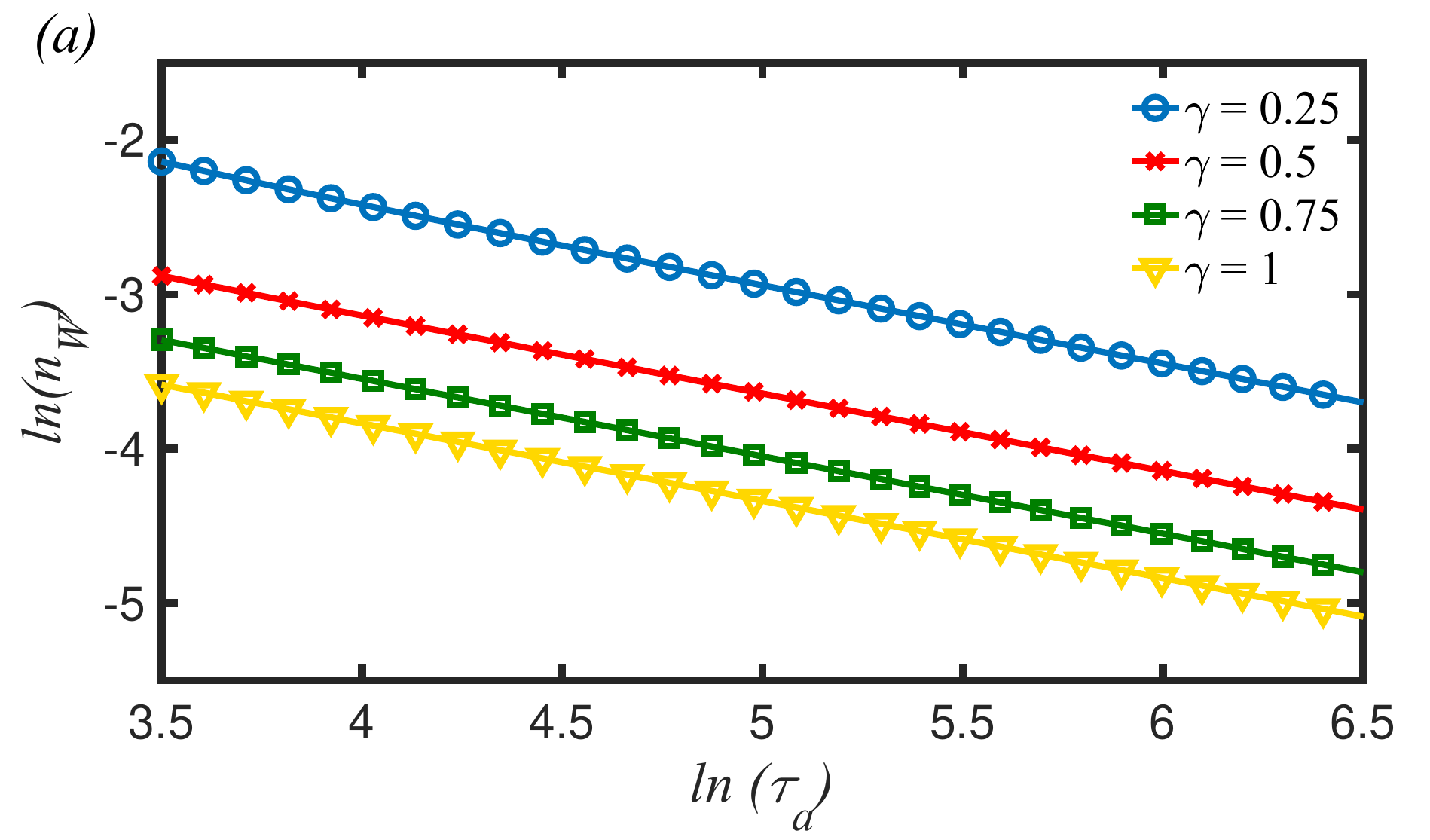}
\includegraphics[width=0.51\linewidth]{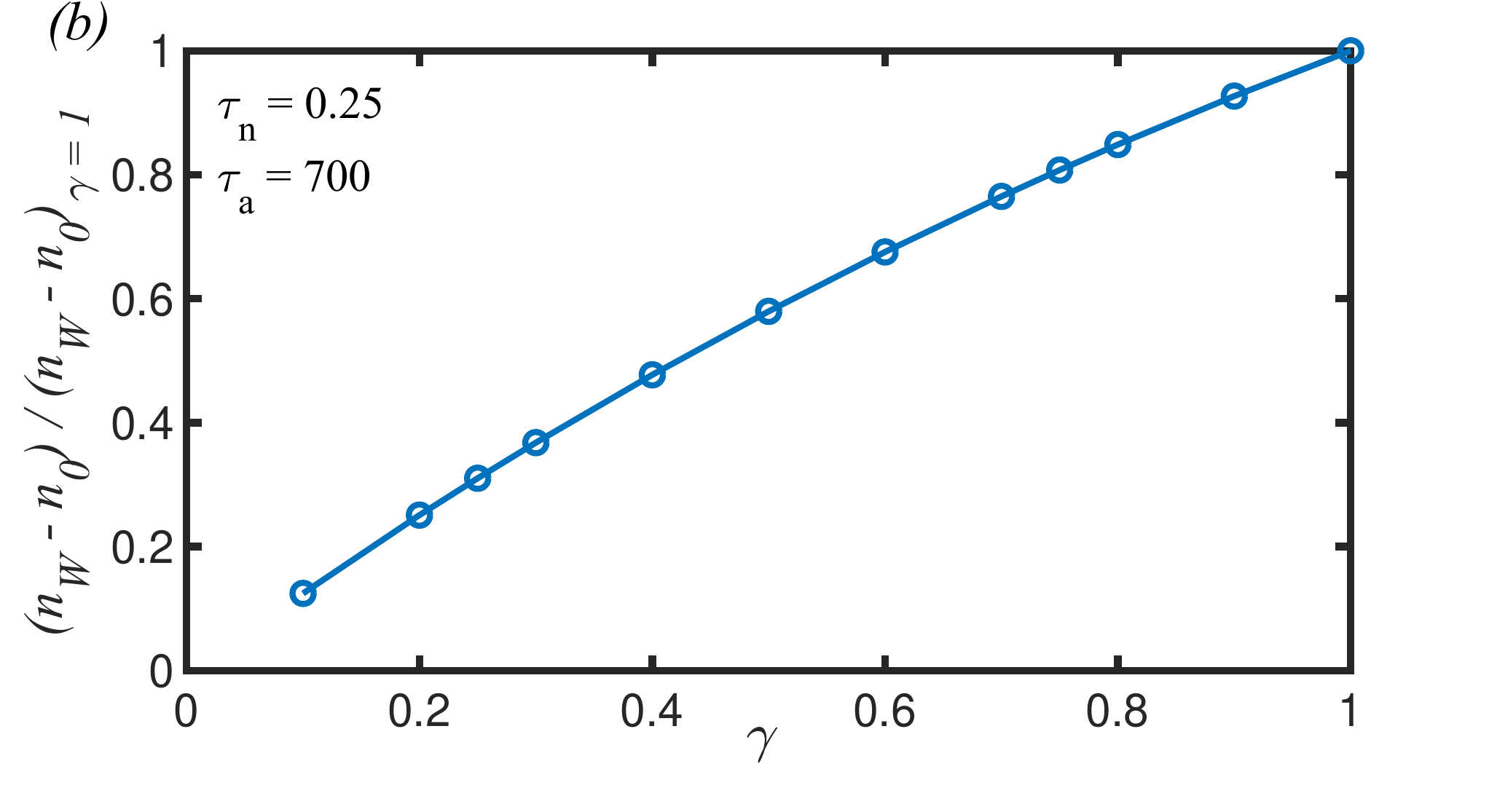}}
\centering		
\caption{(Color online) The effect of anisotropy on the excitation density.
(a) In a noiseless system, $n_W \propto \tau_a^{-0.5}$, which is also proportional to $\gamma^{-1}$. (b) Noise contribution to excitations normalized to its value at $\gamma=1$ increases 
for larger anisotropy values. }
\label{anisotropy}
\end{figure*}
%

%
\begin{figure*}
\centerline{\includegraphics[width=0.5\linewidth]{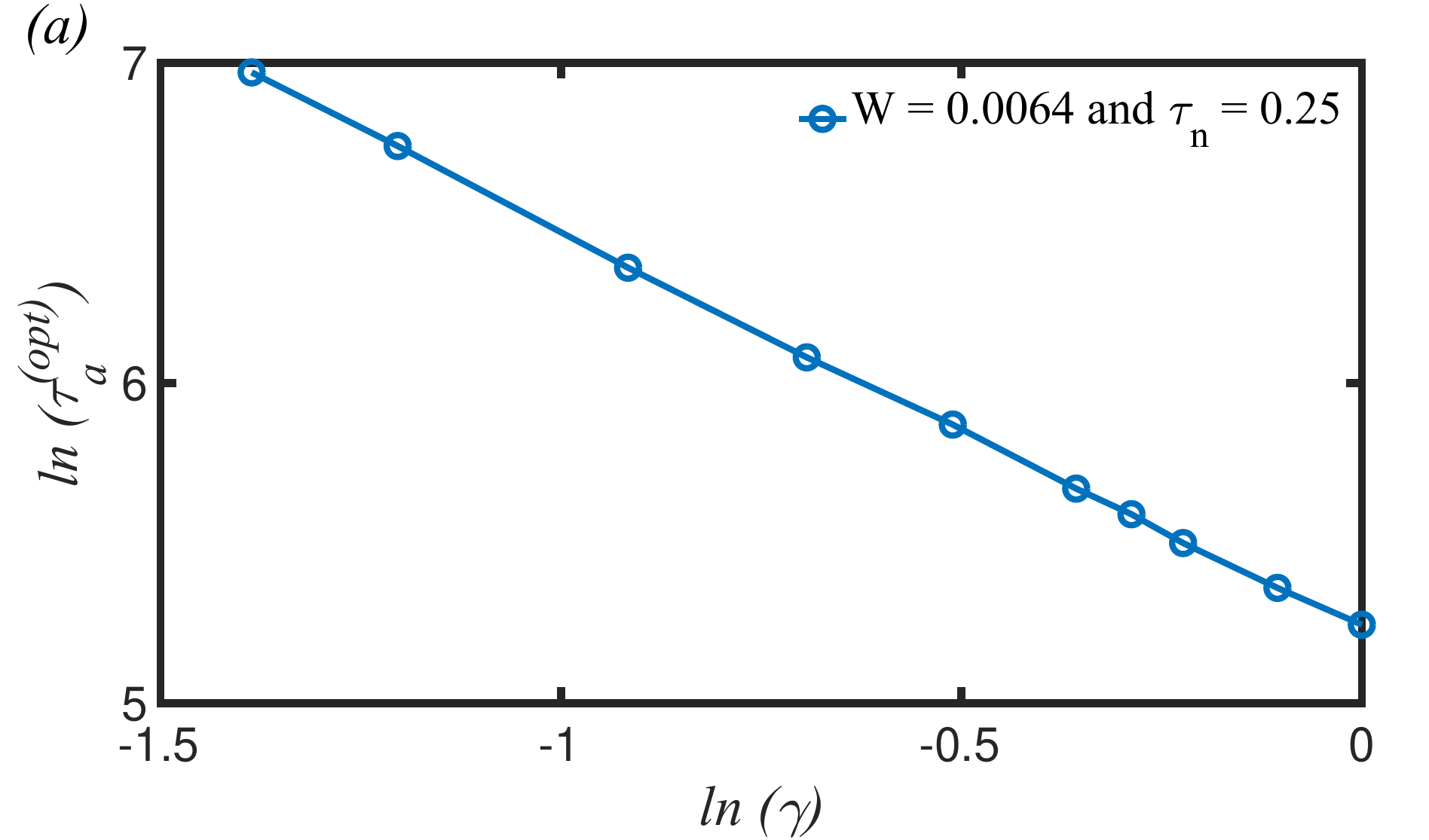}
\includegraphics[width=0.5\linewidth]{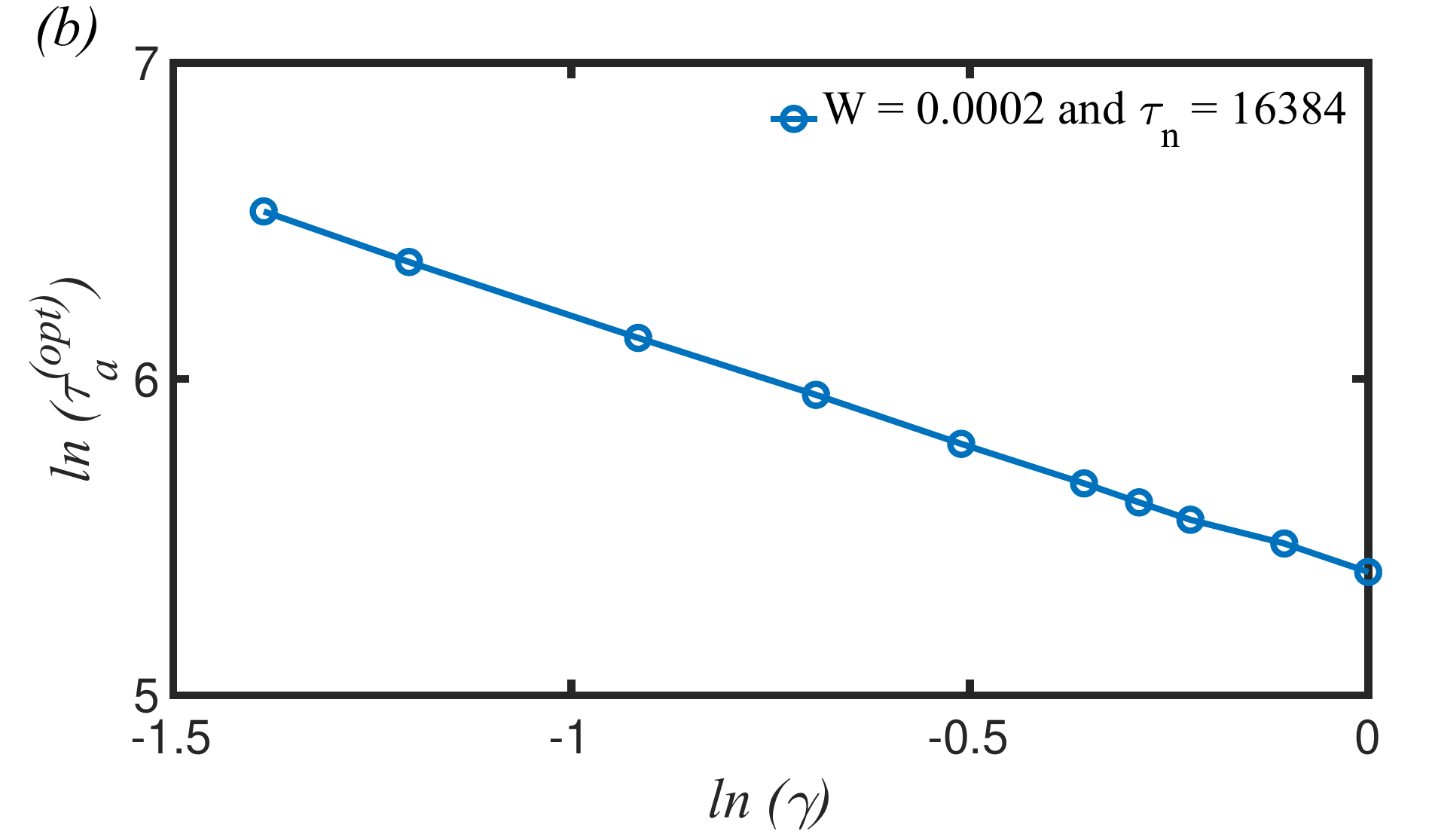}}
\centering 
\caption{(Color online) The optimal annealing time scaling with anisotropy 
coefficient for (a) fast noises, which show the scaling as $\tau^{(opt)}_a \propto\gamma^{-4/3}$ and (b) slow noises, with the scaling behavior of $\tau^{(opt)}_a \propto \gamma^{-0.8}$.}
\label{scaling_gamma}
\end{figure*}
%

%
\begin{equation}
\label{eq:whole}
|\varepsilon_{k}(t_f)\rangle = 
\begin{pmatrix}
\frac{\gamma \sin(k)}{\cos(k) - g_f + \sqrt{(\cos(k)-g_f)^2 + (\gamma \sin (k))^2}} \\
1
\end{pmatrix}.
\end{equation}
%
%
%

The energy gap is reduced by decreasing anisotropy, which consequently leads to a high density of defects.
Fig. \ref{anisotropy}(a) represents the numerical simulation of defects density generation for various anisotropies. As expected, the defects density increases with a reduction in anisotropy.

According to previous studies, in the absence of noise, the density of defects scales with anisotropy as $\gamma^{-1}$ \cite{mukherjee_2007},
%
\begin{equation}
n_{0} = \frac{a}{\gamma} \tau_{a}^{-1/2}. 
\end{equation}
%
However, in the presence of noise, the scaling of defects density with anisotropy is still an open question. 
Figure \ref{anisotropy}(b) depicts the noise contribution to the defects density versus $\gamma$. 
Contrary to the noiseless behavior, the noise induced excitation increases as $\gamma$ grows. Although the noise-induced excitation
does not scale linearly with $\gamma$, our numerical analysis shows that, it scales linearly for small anisotropies.

For fast noises and small anisotropy, the defects density Eq. (\ref{generalnoise}) as a function of anisotropy ($\gamma$) is given by
\begin{equation}
n_W = n_{0} +n _{noise} = \frac{a}{\gamma}  \tau_{a}^{-1/2} + b c \gamma  W^2 \tau_a \tau_n,
\end{equation}
and the optimal annealing time, Eq. (\ref{generaltauopt}), is governed as follows 
\begin{equation}
\tau_{a}^{(opt)} \propto (\gamma^2 W^2 \tau_n)^{-2/3}.
\end{equation}
%
It is clear that, the optimum annealing time indicates power law scaling with anisotropy, i.e., 
$\tau^{(opt)}_a \propto\gamma^{-4/3}$, which is consistent with the numerical simulations displayed in Fig. \ref{scaling_gamma}(a).
 
%
\begin{figure*}
\centerline{\includegraphics[width=0.5\linewidth]{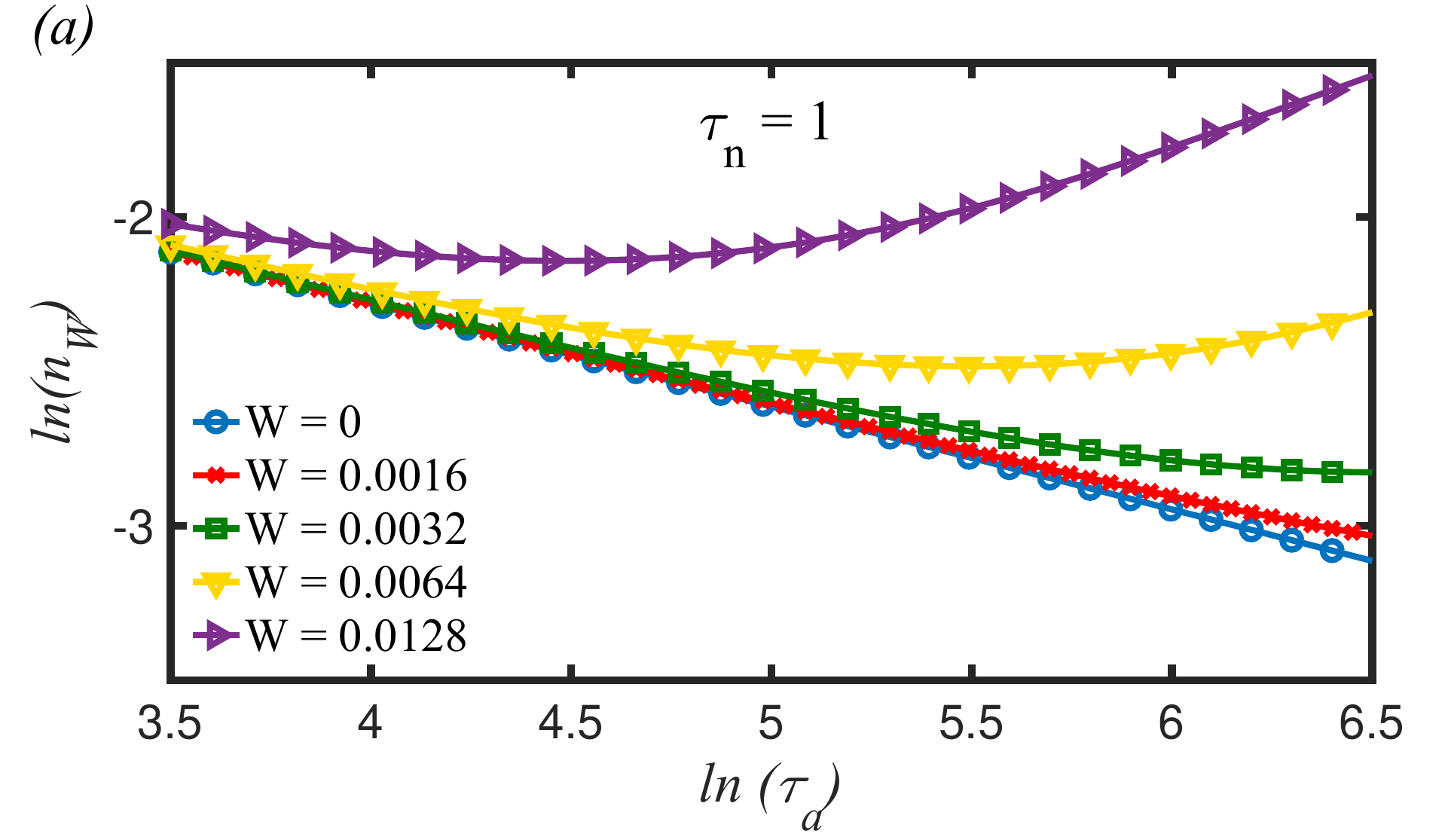}
\includegraphics[width=0.51\linewidth]{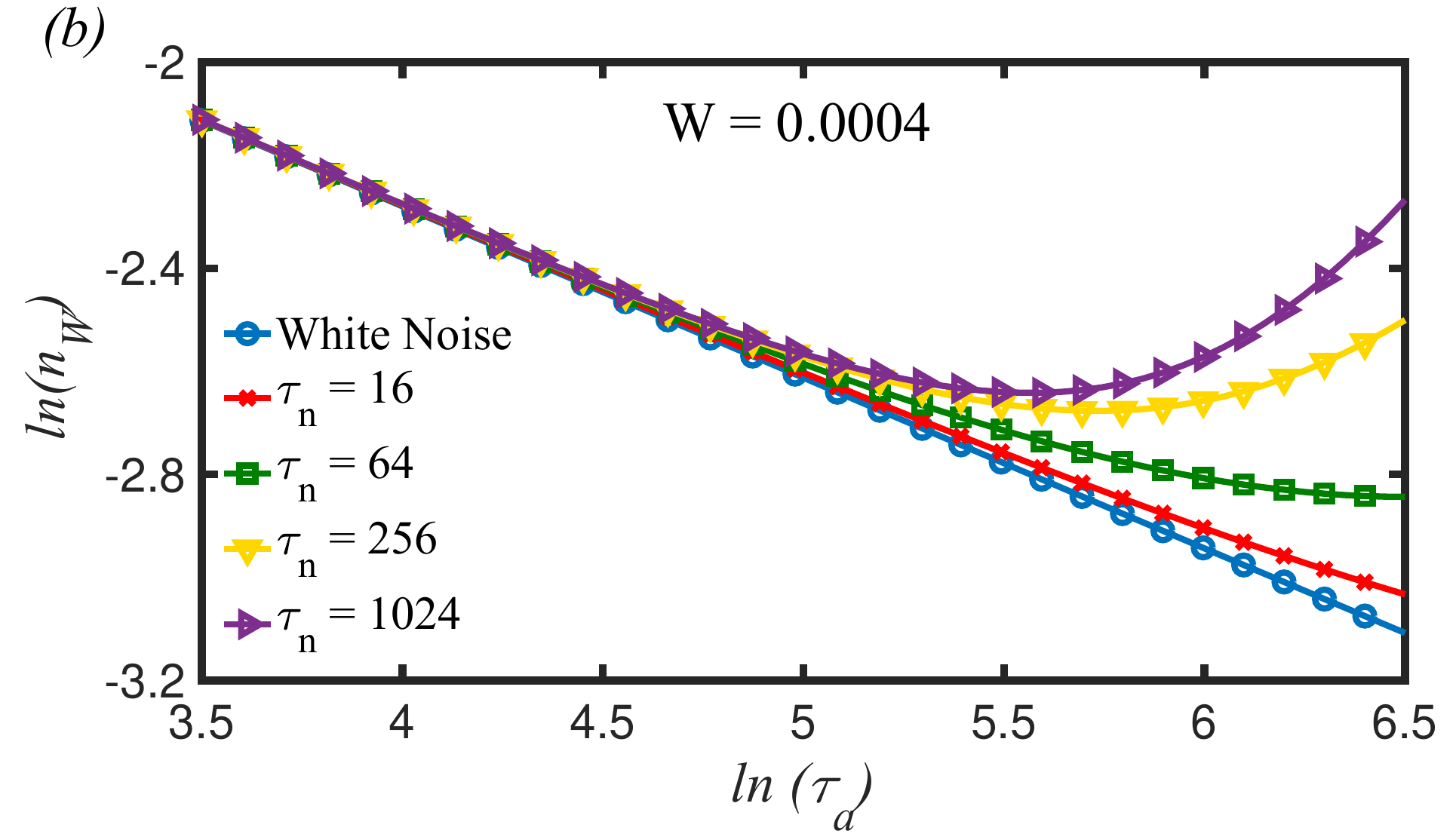}}
\centering 
\caption{(Color online) The excitation density vs. annealing time for quench 
along the gapless line $g_c = -1$  for (a) different powers of noise ($W$) and $\tau_n = 1$   
and (b) $W = 0.0004$ and different values of noise correlation time.}
\label{nex_gamma}
\end{figure*} 
%

For slow noises and small anisotropy, the defect density and optimal annealing time are specified as
%
\begin{equation}
n_W = n_{0} +n _{noise} =  \frac{a}{\gamma} \tau_{a}^{-1/2} + \frac{b \pi \gamma}{2}   W^2 \tau_{a}^{2} \;,  
\end{equation}
%
%
\begin{equation}
\label{eq24}
\frac{d n_W}{d \tau_a} = 0 \Rightarrow \tau_{a}^{(opt)} \propto (\gamma^2 W^2)^{-2/5}.
\end{equation}
%

Fig. \ref{scaling_gamma}(b) depicts the scaling of optimal annealing time with respect to the anisotropy coefficient $\gamma$, for slow noises $\tau_n=16384$. 
The conjecture in Eq. (\ref{eq24}) are confirmed by numerical results presented in Fig. \ref{scaling_gamma}(b) in which the optimal annealing time scales as $\tau^{(opt)}_a \propto \gamma^{-0.8}$.


\section{Linear time dependent anisotropy}\label{sec:level4}

The last part of our investigations is devoted to noisy quench of the anisotropy parameter.
We consider $\gamma(t)=\gamma_0(t)+\eta(t)$ varies in time along the gapless line ($g=-1$), 
in such a way that crosses the multicritical point ($g=-1,\gamma=0$).
The term $\gamma_0(t)$ shows deterministic quench protocol $\gamma_0 (t) = \gamma_i + (\gamma_f - \gamma_i)t/\tau_a$,  which changes 
from $\gamma_i=-2$ at  $t_i = 0$ to the final value of $\gamma_f=2$ at $t_f =\tau_a$ (red dashed-line in Fig. \ref{phase}).
The term $\eta(t)$ is the stochastic part of 
the quenching protocol with zero mean $\langle \eta (t) \rangle = 0$ and two-point correlations given by Eq. (\ref{noise_corr}). 


To implement the exact noise master equation,
the noisy Hamiltonian should be decoupled into noiseless and noisy parts as mentioned earlier in Eq.(\ref{H0H1}). Here, the transverse field is time-independent and 
$\gamma(t)$ is linear time dependent, where the noiseless Hamiltonian $H_0(t)$ is defined as 
$H_0 = \sum_{k>0} \Psi_{k}^{\dagger} \mathcal{H}_{0,k} (t) \Psi_{k}$
with
%
\begin{equation}
\label{eq:whole}
\mathcal{H}_{0,k}(t) = 2(g - \cos(k) ) \hat{\sigma}^{z} +  2 \gamma_0 (t) sin(k) \hat{\sigma}^{x},
\end{equation}
\noindent and the noisy Hamiltonian $H_1$ is given by $H_1 = \sum_{k>0} \Psi_{k}^{\dagger} \mathcal{H}_{1,k} \Psi_{k}$,
where $\mathcal{H}_{1,k} = 2 \sin (k) \hat{\sigma}^{x}$.\\

For the noiseless case ($W=0$), the analytical result \cite{divakaran_defect_2008,divakaran_quenching_2008} 
is justified by the numerical simulations, where the defect density scales as $\tau_a^{\delta}$, with exponent $\delta\approx-0.33$ (Fig. \ref{nex_gamma}(a)).
In the presence of noise, analogous to that of the transverse field quench protocol, the defect density increases by growing 
the noise correlation time and  noise power as shown in Fig. \ref{nex_gamma}(a)-(b).   
These numerical simulations can also be affirmed by implementing the noise power spectrum notion for both fast and slow noises. 
Based on KZM, the quench on anisotropy results in $n_{0} \propto \tau_{a}^{-1/3}$. Hence, Eq. (\ref{noise1}) is modified as the following
%
\begin{equation}
n_W = n_{0} +n _{noise} = a' \tau_{a}^{-1/3} + b' W^2 \tau_{a}^{2} \arctan (\dfrac{c' \tau_n}{\tau_a}),
\label{generalnoise1}
\end{equation}
%
where $a'$, $b'$ and $c'$ are constant coefficients.
Similar to the case of transverse field quench, the constant coefficient $a'$ can be calculated using the LZ formula \cite{divakaran_quenching_2008}
%
\bea\label{LZ3}
p_k = e^{-\pi \tau_Q k^3/2}
\eea
%
which yields,
%
\bea
\label{LZ4}
n_0 =\frac{1}{\pi} \int_{0}^{\pi} p_k dk = \frac{\sqrt[3]{2} \Gamma(1/3) \tau_{Q}^{-1/3}}{3 \pi^{4/3}},
\eea
%
where $\Gamma(x)$ is the Gamma function.
By using the definition $\tau_Q = \tau_a/(\gamma_f-\gamma_i)= \tau_a / 4$, we obtain $a'=\dfrac{ 2\Gamma(1/3) }{3 \pi^{4/3}}$. 
Moreover, the coefficients $b'$ and $c'$ are calculated using the numerical results to be $b'=0.66$ and $c'=3.36$.

\begin{figure*}
\centerline{\includegraphics[width=0.5\linewidth]{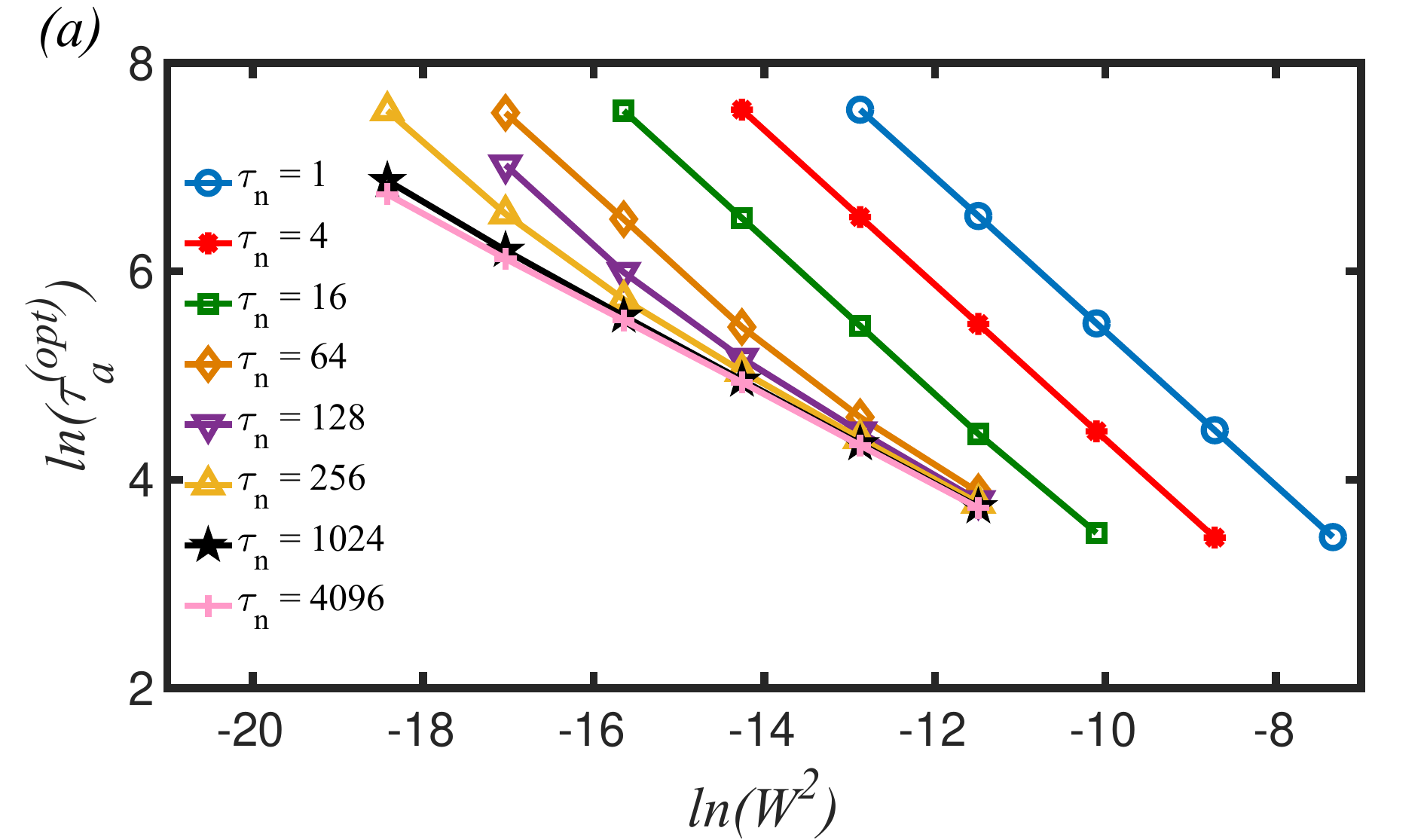}
\includegraphics[width=0.5\linewidth]{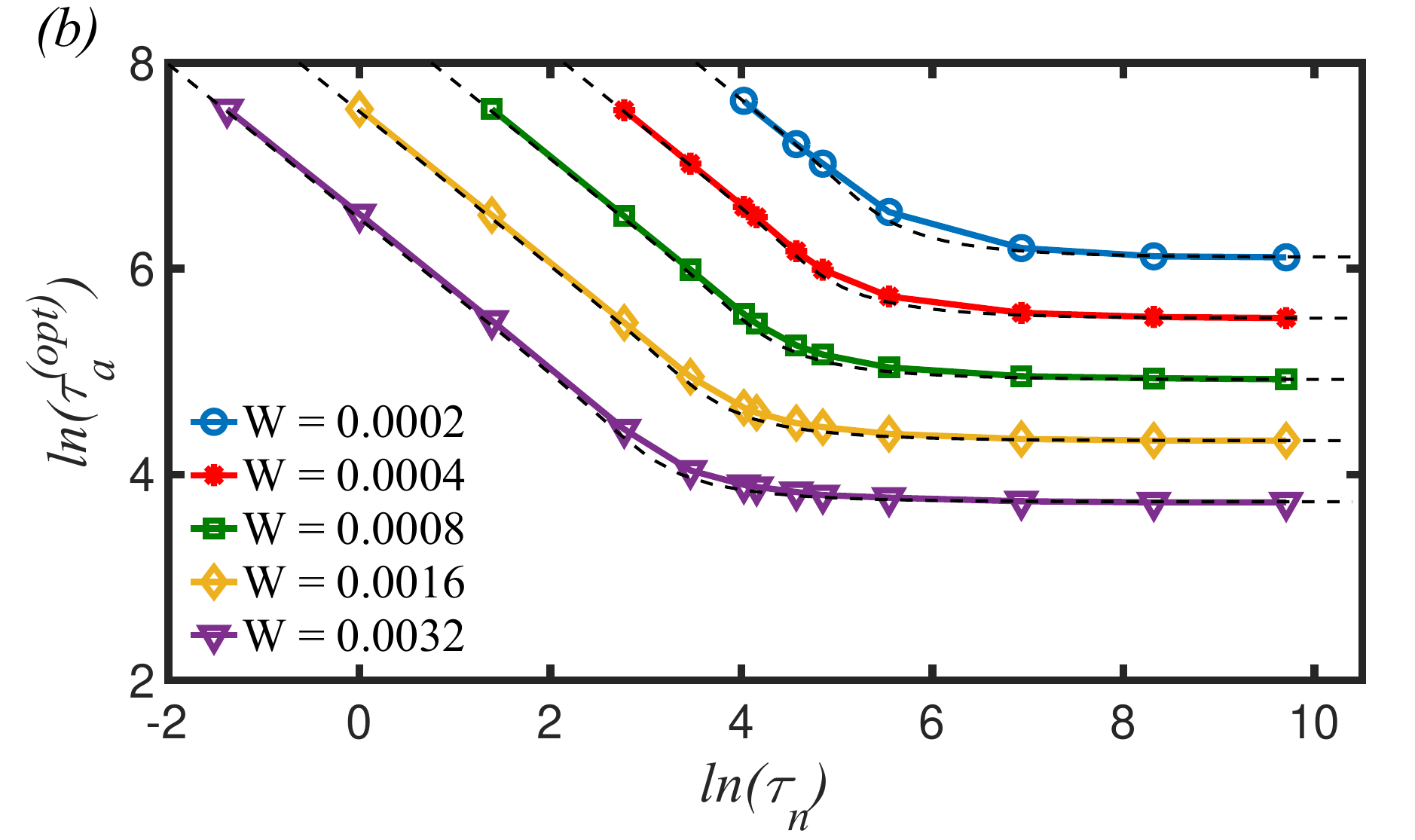}}
\centering 
\caption{(Color online) The scaling of optimal annealing time with (a) 
noise power and (b) noise correlation time. Both for quench along the gapless line ($g=-1$).
The dashed lines are based on the analytic 
conjecture (Eq. \eqref{generalnoise1}) with constants $a' = 2\Gamma(1/3) / 3 \pi^{4/3}$, $b' = 0.66$ and $c' = 3.36$ and the solid lines illustrate our numerical simulations.
}
\label{scaling2}
\end{figure*}

In the fast noise case, Eq. (\ref{generalnoise1}) and the corresponding optimal annealing time 
are reduced to the following forms
%
\begin{equation}
n_W = n_{0} +n _{noise} \simeq a' \tau_{a}^{-1/3} + b' c' W^2 \tau_a \tau_n \;,
\end{equation}
%
%
\begin{equation}
\tau_{a}^{(opt)} \propto (W^2 \tau_n)^{-3/4} \propto (\xi^2)^{-3/4}.
\end{equation}
%
As is clear, the scaling of noise-induced excitation with noise power, noise correlation time and annealing time is the same as that of the transverse filed quench (Eq. (\ref{eq15})). However, the scaling exponent of optimum annealing time
is changed to $\beta=-3/4$. The numerical simulation confirms these finding as shown in Fig. \ref{scaling2} with $\beta\approx-0.75$.

In the slow noise limit, Eq. (\ref{generalnoise1}) and the corresponding optimal annealing time are approximated 
by 
%
\begin{equation}
n_W = n_{0} +n _{noise} \simeq a' \tau_{a}^{-1/3} + \frac{\pi b'}{2} W^2 \tau_{a}^{2},
\end{equation}
%
%
\begin{equation}
\tau_{a}^{(opt)} \propto (W^2)^{-3/7}.
\end{equation}
%
The noise-induced excitation contribution in the defects density, for the slow noise, is the same for both 
anisotropy and transverse field quench. However, the optimum annealing time exponent is $\beta=-3/7$, i.e., $\tau^{(opt)}_a=(W^2)^{-3/7}$,
for linear quenching of anisotropy. 
These result are consistent with our simulations (Fig. \ref{scaling2}) for slow noises, where $\tau^{(opt)}_{a}$  does not change with 
$\tau_n$ and the scaling exponent of optimum annealing time is $\beta\approx-0.43$.

%
%

\section{Conclusions\label{conclusion}}

We have studied the driven dynamics across the critical points of the transverse field quantum $XY$ chain in the presence of 
the {\it colored (correlated) Gaussian noise}. In the noiseless case, universal behaviors of defects density are usually described by the Kibble-Zurek
mechanism, which states that the defects density displays power law scaling in terms of quench time, where the exponent is associated with the critical point. 
However, the mechanism, which causes the topological defects formation, breaks down in the noisy driven control field. 
In other words, in the presence of noise, the large quench time scale provokes more excitations, contrary
to the KZM, where we expect that increasing the quench time scale leads to more adiabatic dynamics.

More precisely, the defect formation is managed by two opposing mechanisms: decreasing the excitation density (adiabaticity) and accumulation
of noise-induced excitation. Consequently, the density of defect exposes a minimum at optimum annealing time due to the competition between these 
two dissenting processes. We have shown that, the defect density formation is affected by the noise correlation time. The longer correlation time enhances the defect density
and the dynamics of defect density changes by increasing the noise correlation time. 

Our results illustrate that, for the noisy driven transverse field through the critical point, the noise-induced defects density scales linearly with 
noise power for both fast/slow noises (small noise correlation time/large noise correlation time). 
The noise-induced defects density scales linearly with annealing time for fast noise, while it scales quadratically with the annealing time for slow noises. Although, the dynamics of defects density is independent of the noise correlation time for slow noises, it shows
linear scaling with noise correlation time for fast noises. 

We have also found that, the optimum driving time scales with the noise power with exponent $\beta=-2/3$, for fast noises and $\beta=-2/5$, 
in the slow noise cases. Furthermore, in contrast to the noiseless cases, the noise-induced excitation increases by enhancing the anisotropy although the band gap becomes larger.

In the case of noisy driven anisotropy through the multicritical point, the dynamics of defects density is the same as that of the driven transverse field.
However, the scaling exponent of optimum annealing time is changed as $\beta=-3/4$, and $\beta=-3/7$, for fast and slow noises, respectively.

Our findings reveal that the time correlation of noise is an important time-scale, which determines the universal dynamics of noisy quenches. This should have a significant impact on the application of quench protocols implemented in the quantum dynamics such as optical lattices or quantum simulations, where noise is an inevitable fluctuation in an experiment.

\appendix

\section{Master equation}\label{APA}

We consider the general Hamiltonian
\begin{equation}
	H = H_0  + \eta (t) H_1  ,
\end{equation}
where $\eta(t)$ is a Gaussian colored noise with zero mean and correlation function:
\begin{equation}
	\langle \eta(t) \eta(t+\tau) \rangle = \frac{\xi^{2}}{2 \tau_{n}}\exp(-|\tau| / \tau_{n}) .
	\label{noise}
\end{equation}
Here, $W^{2} = \dfrac{\xi^{2}}{2 \tau_{n}}$ is the total power of noise and $\tau_{n}$ is  the noise correlation time.

By writing down von Neumann equation
%
\begin{equation}
	\frac{d}{dt} \rho_{\eta}	 (t)  =  -i [ H_0 (t) , \rho_{\eta} (t)]  -i  [ H_1 (t) , \eta (t) \rho_{\eta} (t)].
\end{equation}
%
and taking the ensemble average over all noise realizations
\begin{equation}
	\langle \frac{d}{dt} \rho_{\eta}(t) \rangle =  -i [ H_0 (t) , \langle \rho_{\eta} (t) \rangle]  -i  [ H_1 (t) , \langle \eta (t) \rho_{\eta} (t) \rangle],
	\label{e1}
\end{equation}
we arrive at the following differential equation for the averaged density matrix,  
\begin{equation}
	\frac{d}{dt} \rho (t) =  -i [ H_0 (t) , \rho (t)]  -i  [ H_1 (t) , \langle \eta (t) \rho_{\eta} (t) \rangle],
	\label{e2}
\end{equation}
where $\rho(t) \equiv \langle \rho_{\eta}(t) \rangle$.
Applying Novikov theorem \cite{novikov} for Gaussian noises, we obtain the master equation
\begin{eqnarray}
	\frac{d}{dt}  \rho (t)&=&  -i [ H_0(t) , \rho (t)] \nonumber \\
	& &   -  \frac{\xi^{2}}{2 \tau_{n}} [H_1 , \int_{t_i}^{t} e^{- \frac{|t-s|}  { \tau_{n}}}   [H_1 (s) , \rho (s) ] ds] ,
	\label{master}
\end{eqnarray}
where we define $\Gamma_{k} (t)$ as
\begin{equation}
\Gamma_{k} (t)  =	\int_{t_i}^{t} e^{- \frac{|t-s|}  { \tau_{n}}}   [H_1 (s) , \rho_k (s) ] ds .
\end{equation}

The solution of master equation is obtained by solving the following sets of equations:
	\begin{eqnarray}	
			\frac{d}{dt}  \rho_k (t) = -i [ H_{0,k}(t) , \rho_k (t)]  &-&  \frac{\xi^{2}}{2 \tau_{nj}} [H_1 , \Gamma_{k} (t) ], \nonumber \\
		\frac{d}{dt} \Gamma_{k} (t) = \frac{-\Gamma_{k} (t)}{\tau_{n}} &+&  [H_1 , \rho_{k} (t) ].
	\end{eqnarray}

\section{Transition probability}\label{APB}

In the noiseless case, the transition probability is obtained using the LZ formula. According to LZ formula, when the system crosses the
critical point $g=-1$, only modes close to $k=\pi$ are excited and the modes away from the gap closing mode are remained in the ground state.
 
\begin{figure*}
\centerline{\includegraphics[width=0.5\linewidth]{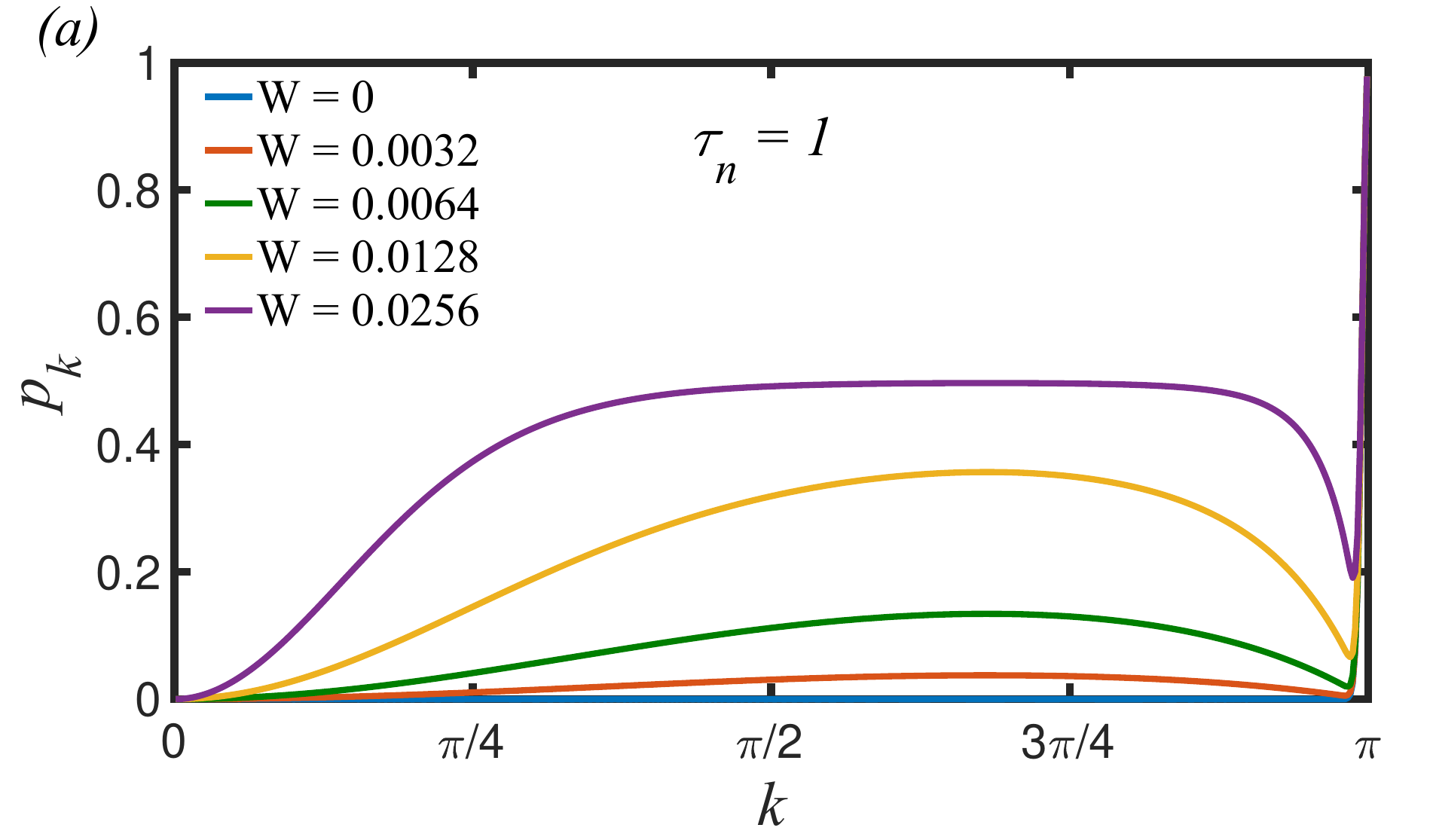}
\includegraphics[width=0.5\linewidth]{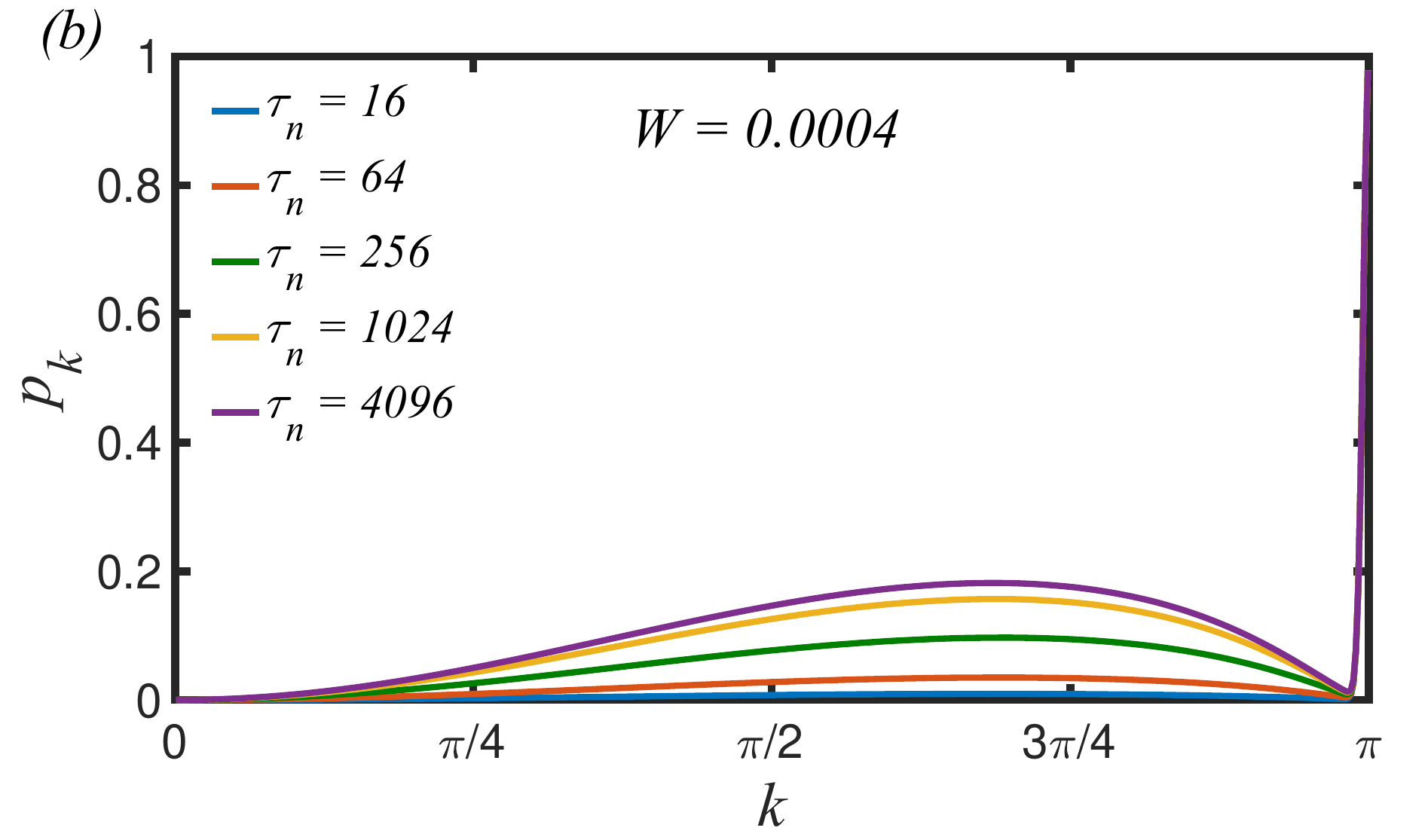}}
\centering 
\caption{(Color online) Excitation probability ($p_k$) for each k-mode of the
transverse field Ising model ($\gamma=1$) for (a) different noise power ($W$) and $\tau_n = 1$,
and (b) $W = 0.0004$ and different values of noise correlation time.}
\label{pk1}
\end{figure*}

In Fig. \ref{pk1} transition probability at the end of quench with $\tau_a = 700$ are depicted for different values of noise power and noise correlation time.
As seen, for $W=0$, the modes close to the gap closing mode are exited $p_{k=\pi}=1$, while the other modes remain at the ground state. 
However, in the presence of noise $W\neq0$, the modes close to the gap closing mode affected less, and noise excites those modes away from the gap closing mode (Fig. \ref{pk1}(a)). 
We should note that the fast noise locks the probability to $1/2$ ($p_k = 0.5$) \cite{Jafari2024}, while in the slow noise case, transition probability oscillates around $p_k = 0.5$ \cite{Jafari2024}. Moreover, increasing the noise correlation time, leads to more excitations (Fig. \ref{pk1}(b)).
%
%

\section{Comparing simulation with the conjecture proposed in Eq. \eqref{generalnoise}}\label{APC}
\begin{figure*}
\centerline{\includegraphics[width=0.5\linewidth]{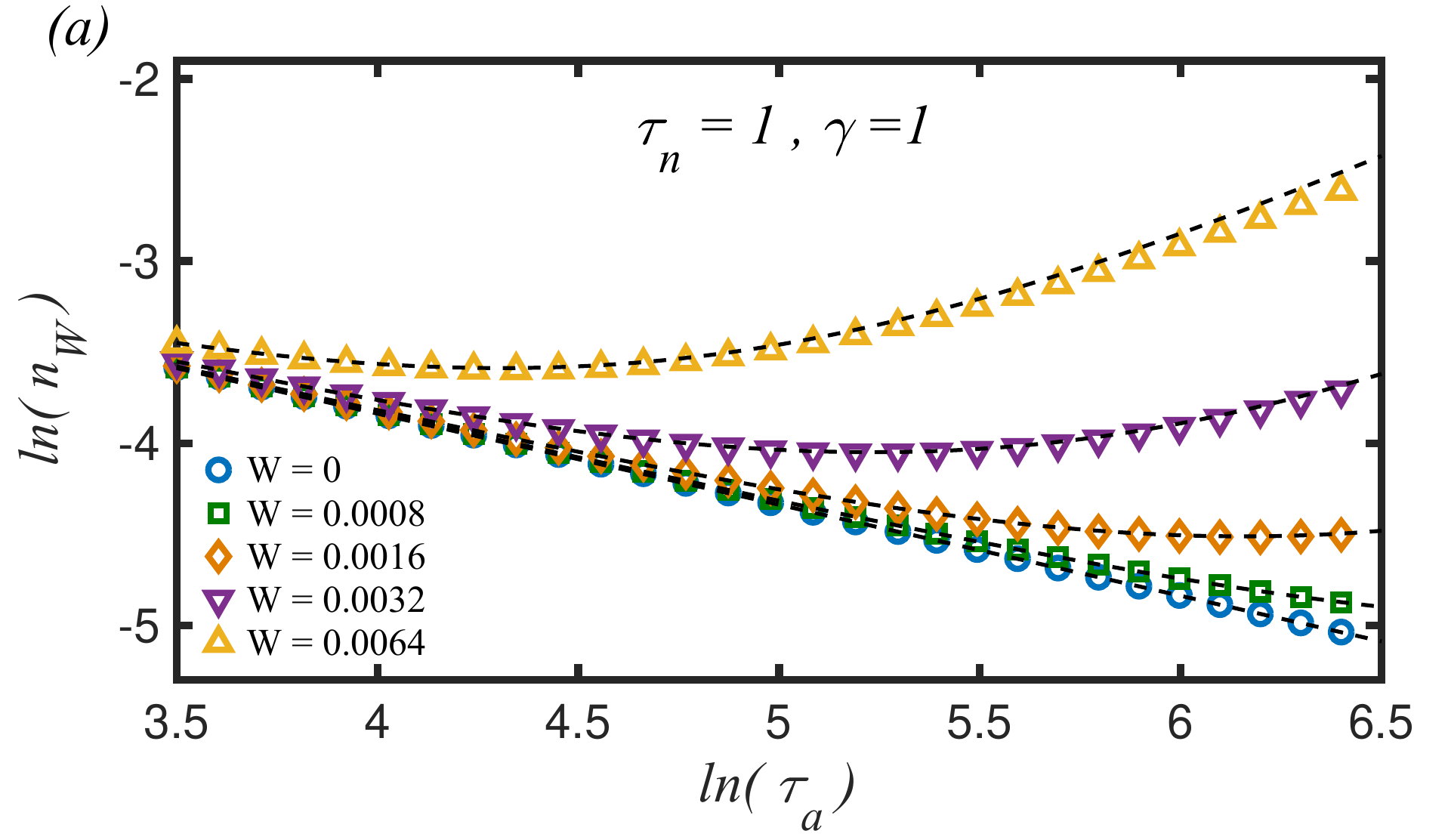}
\includegraphics[width=0.5\linewidth]{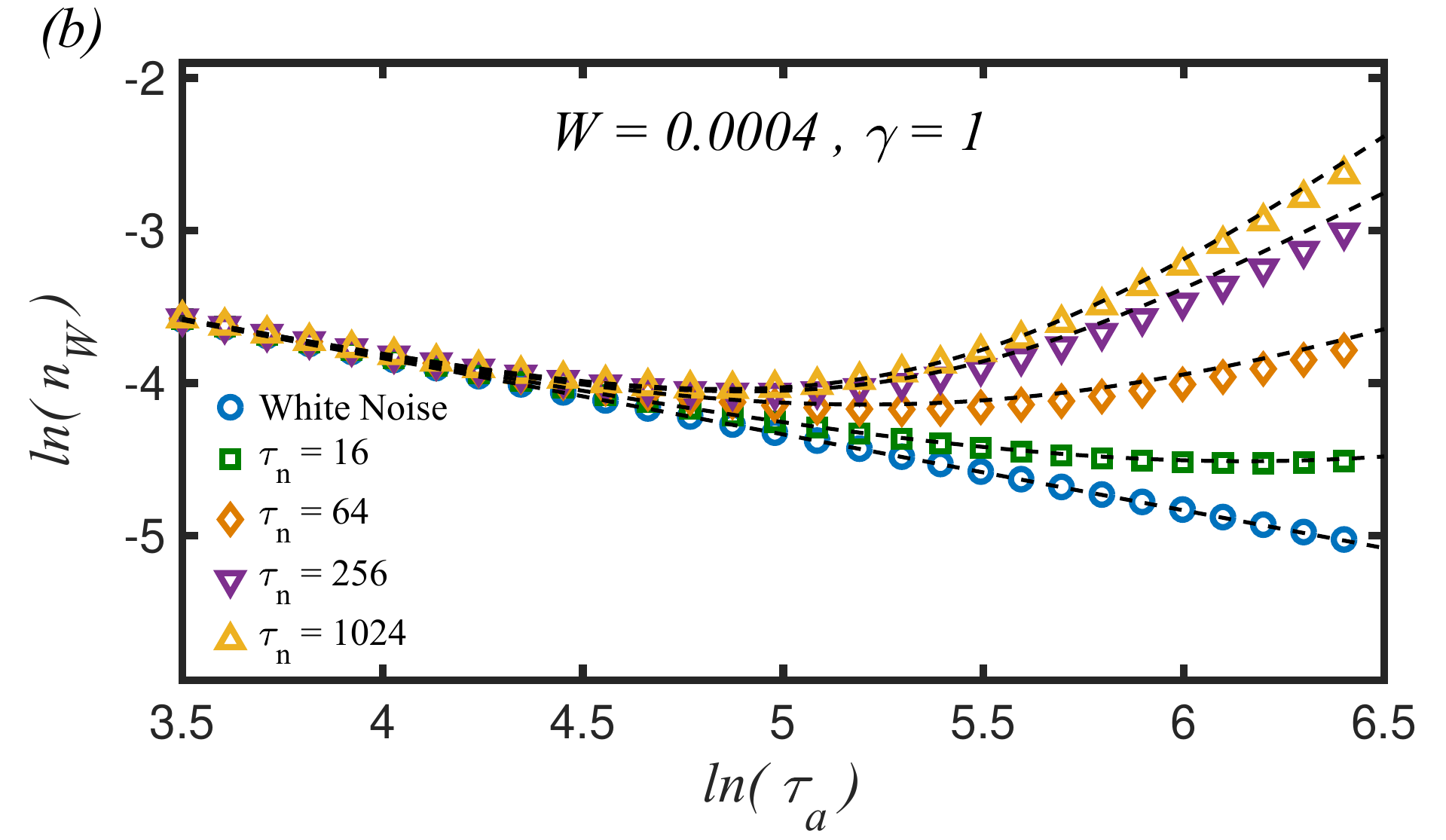}}
\centering 
\caption{(Color online) The density of excitations ($n_W$) vs. annealing time ($\tau_a$) of the
transverse field Ising model ($\gamma=1$) for (a) different noise power ($W$) and $\tau_n = 1$,
and (b) $W = 0.0004$ and different values of noise correlation time. Black dashed lines represents our proposed conjeture in Eq. \eqref{generalnoise}.}
\label{nw_conjecture}
\end{figure*}

In Fig. \ref{scaling}(b), we compared the results obtained from our conjecture in Eqs. \eqref{generalnoise}-\eqref{generaltauopt} with numerical simulations for the optimum annealing time. Here, we compare the aforementioned results for density of excitations at the end of quench.
Fig. \ref{nw_conjecture} represents $\ln(n_W)$ versus $\ln(\tau_a)$ obtained by both numerical simulations and the proposed formula in Eq. \eqref{generalnoise}. As seen, the results based on our conjecture in Eq. \eqref{generalnoise} (the black dashed-lines) show good agreement with the numerical ones. 

%
%


\end{document}